\documentclass[twocolumn,floatfix]{aastex631}

\usepackage[caption=false]{subfig}

\usepackage{float,graphicx,amsmath,tabularx,booktabs,natbib,threeparttable}
\usepackage[figuresleft]{rotating}

\definecolor{dark-red}{rgb}{0.4,0.15,0.15}
\definecolor{dark-blue}{rgb}{0.15,0.15,0.4}
\definecolor{medium-blue}{rgb}{0,0,0.5}
\hypersetup{
    colorlinks, linkcolor={dark-red},
    citecolor={dark-blue}, urlcolor={medium-blue}
}

\newcommand{\beqa}{\begin{eqnarray}} 
\newcommand{\eeqa}{\end{eqnarray}}

\newcommand{\bsub}{\begin{subequations}}
\newcommand{\esub}{\end{subequations}}
\newcommand{\beal}{\begin{align}}
\newcommand{\ealn}{\end{align}}

\newcommand{\kms}{${\rm km~s^{-1}}$}

\newcommand{\Msun}{{\ensuremath{M_{\odot}}}}
\newcommand{\Rsun}{{\ensuremath{R_{\odot}}}}
\newcommand{\mphd}{mag\,100d$^{-1}$\ }

\def \caltech {{Division of Physics, Mathematics and Astronomy, 
California Institute of Technology, Pasadena, CA 91125, USA}}

\def \coo {{Caltech Optical Observatories, California Institute of Technology, Pasadena, CA 91125, USA}}

\def \su {{Department of Astronomy, The Oskar Klein Center, Stockholm University, AlbaNova, 10691 Stockholm, Sweden}}

\begin{document}

\title{\Large{Dramatic rebrightening of the type-changing stripped-envelope supernova SN~2023aew}}

\correspondingauthor{Yashvi Sharma}
\email{yssharma@astro.caltech.edu}

\author[0000-0003-4531-1745]{Yashvi Sharma}
\affiliation{\caltech}

\author[0000-0003-1546-6615]{Jesper Sollerman}
\affiliation{\su} 

\author[0000-0001-5390-8563]{Shrinivas R. Kulkarni}
\affiliation{\caltech} 

\author[0000-0003-1169-1954]{Takashi J. Moriya}
\affiliation{National Astronomical Observatory of Japan, National Institutes of Natural Sciences, 2-21-1 Osawa, Mitaka, Tokyo 181-8588, Japan}
\affiliation{Graduate Institute for Advanced Studies, SOKENDAI, 2-21-1 Osawa, Mitaka, Tokyo 181-8588, Japan}
\affiliation{School of Physics and Astronomy, Monash University, Clayton, Victoria 3800, Australia}

\author[0000-0001-6797-1889]{Steve Schulze}
\affiliation{Center for Interdisciplinary Exploration and Research in Astrophysics (CIERA), Northwestern University, 1800 Sherman Ave., Evanston, IL 60201, USA} 

\author[0000-0003-4800-2737]{Stan Barmentloo}
\affiliation{\su}

\author[0000-0002-9113-7162]{Michael Fausnaugh}
\affiliation{Department of Physics \& Astronomy, Texas Tech University, Lubbock, TX 79410-1051, USA}

\author[0000-0002-3653-5598]{Avishay Gal-Yam}
\affiliation{Department of Particle Physics and Astrophysics, Weizmann Institute of Science, 234 Herzl St, 76100 Rehovot, Israel} 

\author[0000-0001-8005-4030]{Anders Jerkstrand}
\affiliation{\su}

\author[0000-0002-2184-6430]{Tomás Ahumada}
\affiliation{\caltech}

\author[0000-0001-8018-5348]{Eric C. Bellm}
\affiliation{DIRAC Institute, Department of Astronomy, University of Washington, 3910 15th Avenue NE, Seattle, WA 98195, USA}

\author[0000-0001-8372-997X]{Kaustav~K.~Das}
\affiliation{\caltech}

\author{Andrew Drake}
\affiliation{\caltech}

\author[0000-0002-4223-103X]{Christoffer Fremling}
\affiliation{\caltech} 

\author[0000-0002-3841-380X]{Saarah Hall}
\affiliation{Center for Interdisciplinary Exploration and Research in Astrophysics (CIERA), Northwestern University, 1800 Sherman Ave., Evanston, IL 60201, USA}

\author[0000-0002-0129-806X]{K. R. Hinds}
\affiliation{Astrophysics Research Institute, Liverpool John Moores University, Liverpool Science Park, 146 Brownlow Hill, Liverpool L35RF, UK}

\author[0009-0003-6181-4526]{Theophile Jegou du Laz}
\affiliation{\caltech}

\author[0000-0003-2758-159X]{Viraj Karambelkar}
\affiliation{\caltech} 

\author[0000-0002-5619-4938]{Mansi M. Kasliwal}
\affiliation{\caltech}

\author[0000-0002-8532-9395]{Frank J. Masci}
\affiliation{IPAC, California Institute of Technology, 1200 E. California
             Blvd, Pasadena, CA 91125, USA}

\author[0000-0001-9515-478X]{Adam A. Miller}
\affiliation{Center for Interdisciplinary Exploration and Research in Astrophysics (CIERA), Northwestern University, 1800 Sherman Ave., Evanston, IL 60201, USA}

\author[0000-0002-7501-5579]{Guy Nir}
\affiliation{Department of Astronomy, University of California, Berkeley, CA 94720-3411, USA}
             
\author[0000-0001-8472-1996]{Daniel A. Perley}
\affiliation{Astrophysics Research Institute, Liverpool John Moores University, Liverpool Science Park, 146 Brownlow Hill, Liverpool L35RF, UK}

\author[0000-0003-1227-3738]{Josiah N. Purdum}
\affiliation{\coo}

\author[0000-0003-3658-6026]{Yu-Jing Qin}
\affiliation{\caltech}

\author[0000-0002-5683-2389]{Nabeel Rehemtulla}
\affiliation{Department of Physics and Astronomy, Northwestern University, 2145 Sheridan Road, Evanston, IL 60208, USA}
\affiliation{Center for Interdisciplinary Exploration and Research in Astrophysics (CIERA), 1800 Sherman Ave., Evanston, IL 60201, USA}

\author[0000-0003-0427-8387]{R. Michael Rich}
\affiliation{University of California Los Angeles, Department of Physics \& Astronomy, Los Angeles, CA, USA}

\author[0000-0002-0387-370X]{Reed L. Riddle}
\affiliation{\coo}

\author[0000-0003-4189-9668]{Antonio C. Rodriguez}
\affiliation{\caltech}

\author[0000-0003-4725-4481]{Sam Rose}
\affiliation{\caltech}

\author[0000-0001-8426-5732]{Jean Somalwar}
\affiliation{\caltech}

\author[0000-0003-0733-2916]{Jacob L. Wise}
\affiliation{Astrophysics Research Institute, Liverpool John Moores University, Liverpool Science Park, 146 Brownlow Hill, Liverpool L35RF, UK}

\author[0000-0002-9998-6732]{Avery Wold}
\affiliation{IPAC, California Institute of Technology, 1200 E. California
             Blvd, Pasadena, CA 91125, USA}

\author[0000-0003-1710-9339]{Lin Yan}
\affiliation{\caltech} 

\author[0000-0001-6747-8509]{Yuhan Yao}
\affiliation{Miller Institute for Basic Research in Science, 468 Donner Lab, Berkeley, CA 94720, USA}
\affiliation{Department of Astronomy, University of California, Berkeley, CA 94720, USA}


\begin{abstract}

Multi-peaked supernovae with precursors, dramatic light-curve rebrightenings, and spectral transformation are rare, but are being discovered in increasing numbers by modern night-sky transient surveys like the Zwicky Transient Facility (ZTF). Here, we present the observations and analysis of SN~2023aew, which showed a dramatic increase in brightness following an initial luminous ($-17.4$ mag) and long ($\sim$100 days) unusual first peak (possibly precursor). SN~2023aew was classified as a Type IIb supernova during the first peak but changed its type to resemble a stripped-envelope supernova (SESN) after the marked rebrightening. We present comparisons of SN~2023aew's spectral evolution with SESN subtypes and argue that it is similar to SNe Ibc during its main peak. P-Cygni Balmer lines are present during the first peak, but vanish during the second peak's photospheric phase, before H$\alpha$ resurfaces again during the nebular phase. The nebular lines ([\ion{O}{1}], [\ion{Ca}{2}], \ion{Mg}{1}], H$\alpha$) exhibit a double-peaked structure which hints towards a clumpy or non-spherical ejecta. We analyze the second peak in the light curve of SN~2023aew and find it to be broader than normal SESNe as well as requiring a very high $^{56}$Ni mass to power the peak luminosity. We discuss the possible origins of SN~2023aew including an eruption scenario where a part of the envelope is ejected during the first peak which also powers the second peak of the light curve through SN-CSM interaction.

\end{abstract}


\section{Introduction} \label{sec:intro}

Core-collapse (CC) supernovae (SNe) mark the final explosions of massive stars ($\gtrsim8~M_\odot$), and stripped-envelope SNe (SESNe) represent CC in stars that have lost most - or all - of their envelopes prior to explosion. This includes Type IIb SNe (some H left), SNe Ib (no H, some He), and SNe Ic (neither H nor He, \citealp{Gal-Yam2017}). 

Even though we now have hundreds of well-observed SESNe, there are still several open questions regarding their nature, both when it comes to their progenitor stars and their powering mechanism. Binarity seems to be a key component for stripping their envelopes, with arguments supported by relatively low ejecta masses, large relative rates \citep{Smith2011}, and direct evidence of a binary system after the SESN 2022jli \citep{Ping2023}. These deduced ejecta masses often come from comparisons with simple analytical models \cite[e.g.,][]{arnett1982,YangHaffet,Barbarino2021}, that match reasonably well with the observed light curves assuming powering by radioactive $^{56}$Ni. However, modern explosion models are unable to produce the amount of radioactive nickel required for the brighter Type Ibc SNe \citep{Sollerman2022} and some SESNe show light-curve features that are not compatible with the standard scenario. Such unusual SESNe have emerged from the large samples of SNe now available, and include double bump light curves (LCs) for example for SN 2019cad \citep{Claudia2021}, SN 2022xxf \citep{Hanin2023} and SN 2022jli \citep{Ping2023}, where different powering mechanisms were suggested in each of these cases for explaining the second LC bump. 

In this paper we present the unusual stripped-envelope SN\,2023aew (ZTF23aaawbsc) discovered as part of the Zwicky Transient Facility Bright Transient Survey \cite[BTS;][]{Fremling2020,Perley2020,Rehemtulla2024}. This supernova shows an unprecedented first peak with a broad light curve and a slight plateau followed by another unusually broad second peak light curve. Spectrally, this object is clearly a SESN, but unlike any previous such objects.

The paper is organized as follows. In \S\ref{sec:obs} we present the discovery and the observations of our SN, as well as details about the data reductions and calibrations. Section \S\ref{sec:analysis} presents an analysis of the photometric and spectroscopic data as well as comparisons to a number of similar SNe from the literature. In \S\ref{sec:discussion} we discuss in particular the mechanisms that could power the main light curve peak of SN 2023aew, and in this connection we also present a few other objects with relevant observations. Finally, \S\ref{sec:conclusion} presents our conclusions and a short discussion where we put our results in context.

\section{Observations} \label{sec:obs}
In this section, we present our observations of SN~2023aew obtained over 300 days with multiple instruments and describe the data processing methods.

\subsection{Discovery}
\begin{figure*}[htbp]
    \centering
    \includegraphics[width=0.95\textwidth]{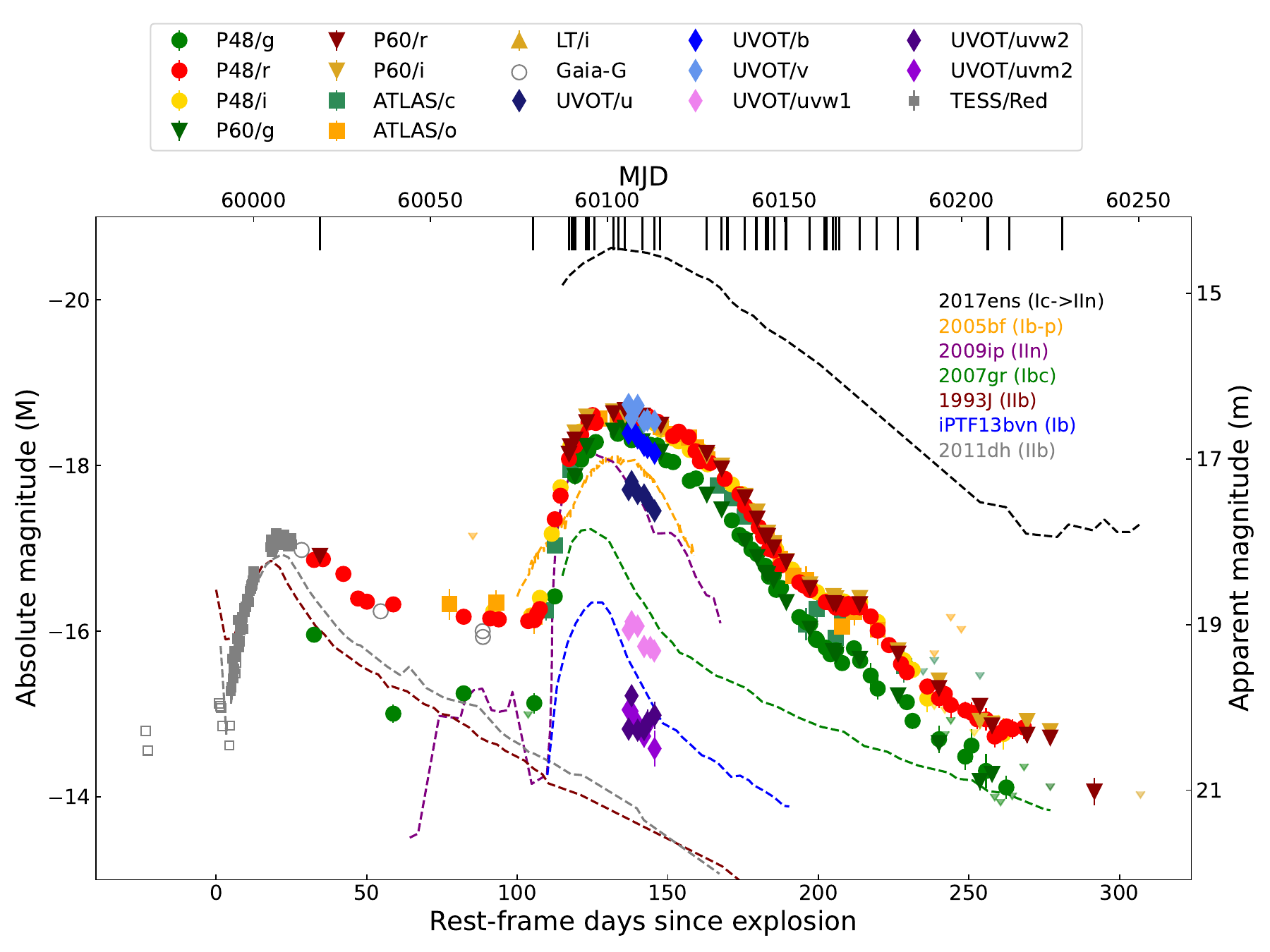}
    \caption{Light curve of SN~2023aew. The TESS observations cover the rise of the first peak and are shown in gray squares, with 5$\sigma$ significance denoted with full markers and 3$\sigma$ with empty markers. The spectroscopic epochs are marked with black lines at the top. Also shown for comparison are absolute magnitude $r$-band light curves of SNe 1993J (IIb), 2011dh (IIb), 2005bf (Ib-pec), 2007gr (Ic), 2009ip (IIn), iPTF13bvn (Ib), and the type changing 2017ens (Ic to IIn). SNe 1993J and 2011dh are shifted to match their first detection with SN~2023aew explosion epoch. Other comparison SNe are shifted by $\sim$100 days to match the start of the second peak. The overall light curve shape is somewhat similar to that of SN~2009ip but is broader, with the second peak having similar broadness as SN~2017ens.}
    \label{fig:lc}
\end{figure*}

SN~2023aew was detected in Zwicky Transient Facility (ZTF; \citealt{Bellm2019b,graham2019,Dekany20}) data obtained with the Palomar Schmidt 48-inch Samuel Oschin telescope (P48), on 2023-01-23 (MJD 59967.511) and the discovery was reported to the Transient Name Server (TNS\footnote{\url{https://www.wis-tns.org}}) by ALeRCE \citep{alerce, aew_discovery_tns}. This first ZTF detection magnitude was 18.05 in the $r$ band at the J2000.0 coordinates $\alpha = 17^h40^m51.395^s$, $\delta = +66^{\circ}12\arcmin22\farcs62$. The transient is apparently positioned in the outskirts of the spiral host galaxy SDSS J174050.55+661220.7. The transient was subsequently reported to TNS by Gaia \citep{gaia} in February, ATLAS \citep{Tonry2018} in March, and by MASTER \citep{master} in May when it began to brighten again. Gaia reported an 18.16 mag detection in $G$-Gaia band two days before the ZTF discovery (i.e. at MJD 59965.284). The last $3\sigma$ upper limit is $\sim200$ days before first detection in ATLAS $o$ band and $\sim500$ days before first detection in ZTF $r$ band. However, the Transiting Exoplanet Survey Satellite (TESS; \citealp{tess}) had a serendipitous two months of coverage of SN~2023aew right before the ZTF discovery (from MJD 59910 to 59962). TESS captured a slow, 30~day rise of this SN where ground-based telescopes only caught the tail of this transient at its ZTF discovery (see Figure~\ref{fig:lc}). We derive an explosion epoch of MJD 59936.18 $\pm$ 1.4 days by fitting a power law to an 8-hour binned TESS-Red band light curve (details in \S\ref{sec:tess}, \citealp{Fausnaugh_2021}). Therefore, all phases throughout this paper will be reported with respect to this explosion epoch estimate.

The transient was spectroscopically classified as a SN IIb by ZTF \citep{aew_classification_tns} on 2023-01-27 (4 days after ZTF discovery) with a spectrum obtained using the Spectral Energy Distribution Machine (SEDM; \citealp{Ben-Ami12,sedm2018}) on the Palomar 60-inch telescope (P60; \citealp{Cenko2006}) and its \texttt{superfit} \citep{superfit} match to SN IIb templates at a redshift of $z = 0.025$. This is consistent with the redshift of $z = 0.0255 \pm 0.0001$ obtained from the narrowest lines in our late high signal-to-noise ratio Keck2/ESI spectrum (\S2.3). Using a flat cosmology with 
H$_{0} = 70$~\kms and $\Omega_m = 0.3$, 
this redshift corresponds to a luminosity distance of 111 Mpc. The transient, initially classified as SN IIb was observed to have a smooth initial light-curve decline for $\sim$25 days which then turned into a slow plateau for another $\sim50$ days in ground-based optical data. 
The great surprise came with the rapid rebrightening of SN~2023aew, which started around 2023-04-11 (MJD 60045), wherein the SN rose by $\sim2$ magnitudes in $\sim10$ days. As spectroscopic campaigns began in earnest, \citet{aew_astronote1} reported their spectrum taken on 2023-04-20 with SPRAT \citep{SPRAT} on the Liverpool Telescope \citep{lt} to be consistent with SN Ib templates albeit at a slightly higher redshift using SNID \citep{snid2007}. \citet{aew_astronote2} reported three more spectra taken with SNIFS on the University of Hawaii 2.2m telescope on 2023-04-23, 2023-04-25, and 2023-04-29 which were more consistent with late-time SN Ic templates. 

\subsection{Optical photometry}

We obtained forced PSF photometry via the ZTF forced photometry service \citep{Masci2019, ztfipac_doi} in $g$, $r$, and $i$ bands and via ATLAS forced photometry service \citep{Tonry2018, atlassmith2020} in $c$ and $o$ bands. Additional photometry was obtained with the Rainbow camera on P60 and processed with the automatic image subtraction pipeline \texttt{FPipe} \citep{Fremling2016}. The Gaia-G band photometry was obtained from Gaia Alerts service\footnote{\url{http://gsaweb.ast.cam.ac.uk/alerts/home}}. All photometry is corrected for Milky Way extinction using the Python package \texttt{extinction} \citep{extinction2016}, the dust extinction law from \citet{fitzp1999}, the \citet{Schlafly_2011} dust map, \textit{E}($B-V$) = 0.0386 mag and an R$_V$ of 3.1. All measurements are converted into flux units for the analysis. We do not account for the host reddening given the transient is at the outskirts of the host galaxy.

\subsection{TESS photometry}\label{sec:tess}
SN~2023aew had serendipitous coverage from TESS observations of Sector 59 and 60 from MJD 59910 to 59962 -- 2 months before the first ZTF detection. The TESS-Red\footnote{\url{http://svo2.cab.inta-csic.es/theory/fps/index.php?mode=browse&gname=TESS&asttype=}} filter extends from 5802.57~\AA\ to 11171.45~\AA\ with a reference wavelength of 7697.60~\AA. Image subtraction and forced photometry at SN~2023aew's location was carried out according to the methodology in \citet{Fausnaugh_2021}. The differential flux has a cadence of 200 seconds but was binned into 6-hour bins and converted into Vega magnitudes. These were further converted to the AB system and corrected for MW extinction following the method in previous section and using the TESS-Red reference wavelength. The binned TESS photometry is included in Table~\ref{tab:tess}.

\subsection{\textit{Swift} Ultraviolet/Optical telescope photometry}

The field was observed with the Ultraviolet/Optical Telescope \citep{Roming2005a} (UVOT) aboard the \textit{Swift} satellite \citep{Gehrels2004a} between MJD$=$60076.47 and 60085.44 in $w2$, $m2$, $w1$, $u$, $b$, $v$. The science-ready data was retrieved from the \textit{Swift} archive\footnote{\href{https://www.swift.ac.uk/swift_portal}{https://www.swift.ac.uk/swift\_portal}}. In December 2023, deep template images were obtained in all filters to remove the host contamination from the transient photometry. Then 
all sky exposures for a given epoch and filter were co-added to boost the signal-to-noise ratio using \texttt{uvotimsum} in HEAsoft\footnote{\href{https://heasarc.gsfc.nasa.gov/docs/software/heasoft/}{https://heasarc.gsfc.nasa.gov/docs/software/heasoft/}} version 6.31.1. Afterwards, the brightness of the SN was measured with the \textit{Swift} tool {\tt uvotsource}. The source aperture had a radius of $5''$, while the background region had a significantly larger radius. We measured the host contribution from the December 2023 templates using the same source and background apertures and subtracted this contribution from the transient flux measurements. All measurements were calibrated with the latest calibration files from November 2021 and converted to the AB system following \citet{Breeveld2011a}. Table \ref{tab:app:xrt} summarises all measurements (not corrected for reddening).

\subsection{\textit{Swift} X-ray telescope measurements}

While monitoring SN~2023aew with UVOT between MJD$=$60076.47 and 60085.44, \textit{Swift} also observed the field with its onboard X-ray telescope XRT between 0.3 and 10 keV in photon-counting mode \citep{Burrows2005a}. This data was analyzed with the online tools of the UK \textit{Swift} team\footnote{\href{https://www.swift.ac.uk/user_objects/}{https://www.swift.ac.uk/user\_objects}} that use the software package HEASoft version 6.32 and methods described in \citet{Evans2007a, Evans2009a}.

SN~2023aew evaded detection in all epochs. The median $3\sigma$ count-rate limit of each observing block is $8\times10^{-3}~\rm s^{-1}$ (0.3--10~keV). Coadding all data pushes the $3\sigma$ count-rate limits to $1.4\times10^{-3}~\rm s^{-1}$. A list of the limits from the stacking analysis is shown in Table~\ref{tab:xray}. To convert the count-rate limits into a flux, a power-law spectrum was assumed with a photon index\footnote{The photon index is defined as the power-law index of the photon flux density ($N(E)\propto E^{-\Gamma}$).} of $\Gamma=2$ and a Galactic neutral hydrogen column density of $3.4\times10^{20}$~cm$^{-2}$ \citep{HI4PI2016a}. The coadded count-rate limit corresponds to an unabsorbed flux of $<5.5\times10^{-14}~{\rm erg\,cm}^{-2}\,{\rm s}^{-1}$ between 0.3--10 keV and luminosity of $<7.8\times10^{40}~{\rm erg\,s}^{-1}$. The flux and luminosity limits of the individual bins are shown in Table~\ref{tab:xray}.

\subsection{Optical spectroscopy}
\begin{figure*}
    \centering
    \includegraphics[width=1.25\textwidth,angle=90,origin=c]{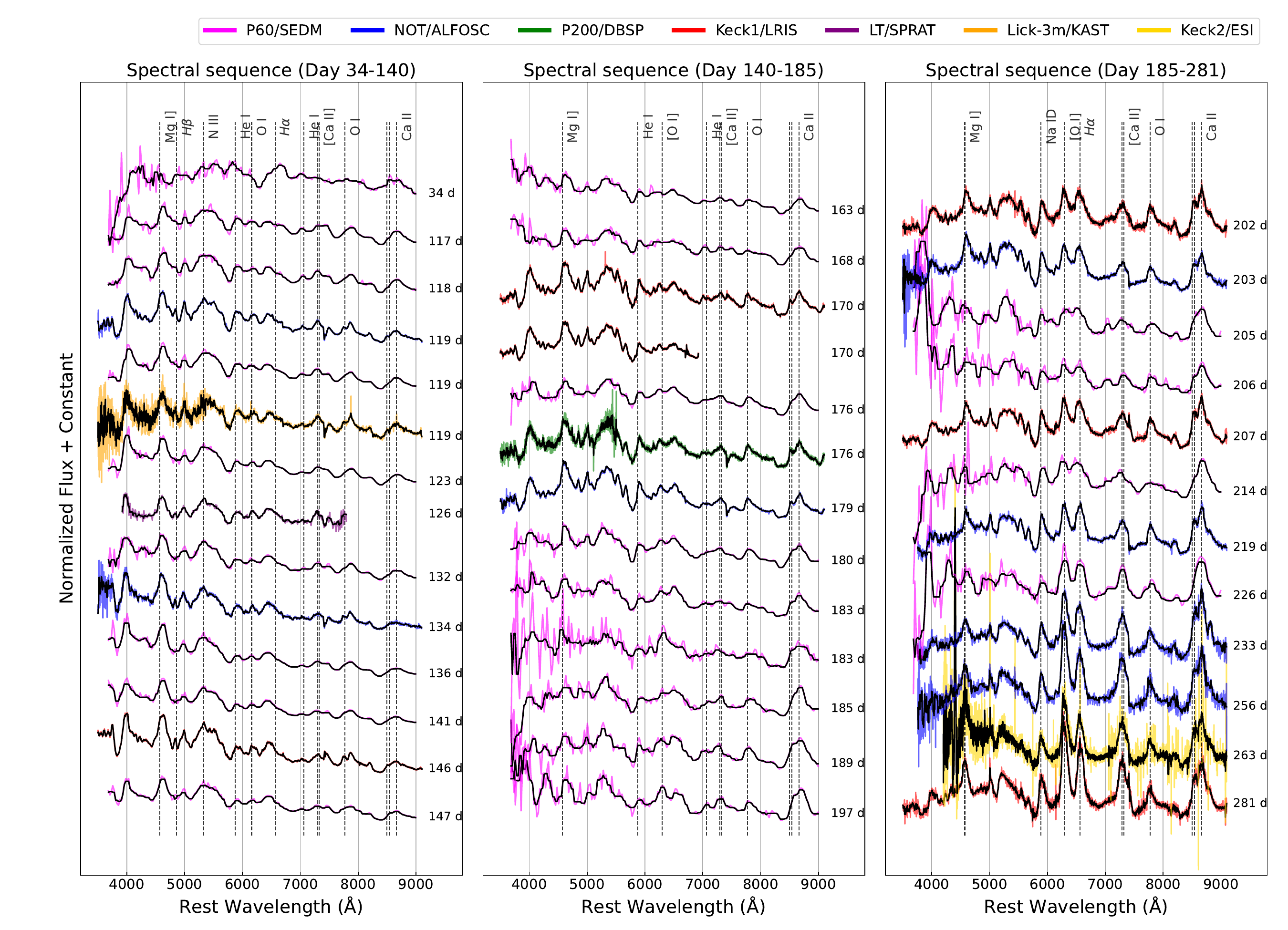}
    \caption{Spectral sequence of SN~2023aew from day 34 to day 281. The black lines are smoothened spectra (using median filter) and the colored lines are original spectra, with different colors depicting different instruments. Some important spectral lines are marked with gray dashed lines. In left panel, the Balmer lines are visible in the first spectrum at 34 days, then H$\alpha$ seemingly shows up again in the nebular phase spectra in the third panel.}
    \label{fig:specseq}
\end{figure*}

We obtained a comprehensive spectroscopic follow-up dataset from many facilities at a variety of spectral resolutions (e.g., KeckI/LRIS $R\sim800$\,--$1400$, P200/DBSP $R\sim1000$, NOT/ALFOSC $R\sim360$, P60/SEDM $R\sim100$) to study the evolution of this SN. In total, we have 41 spectra covering epochs from 34 days to 281 days since explosion. Table~\ref{tab:inst} lists the facilities, instruments, and data processing software references. The spectral sequence is listed in Table~\ref{tab:speclist} and shown in Figure~\ref{fig:specseq}. All the spectra were corrected for Milky Way extinction using the same procedure as for the photometry, then scaled to match the synthetic photometry from the spectra with the contemporaneous host-subtracted ZTF $r$-band data. The SN redshift ($z = 0.0255 \pm 0.0001$) was obtained from the narrowest lines in our highest resolution Keck2/ESI spectrum in the absence of a pre-existing host redshift measurement. The spectra will be made available on WISeREP\footnote{https://www.wiserep.org/} \citep{wiserep}.

\begin{table}
    \centering
    \caption{Description of spectrographs used for follow-up and the corresponding data reduction pipelines}
    \footnotesize
    \begin{threeparttable}
    \begin{tabularx}{0.49\textwidth}{ccc}
    \toprule
    \toprule
        \textbf{Instrument} & \textbf{Telescope} & \textbf{Software} \\
        \midrule
        SEDM\tnote{1} & Palomar 60-inch & \texttt{pySEDM}\tnote{2} \\
        ALFOSC\tnote{3} & Nordic Optical Telescope & PyNOT\tnote{4}, \texttt{PypeIt} \\
        DBSP\tnote{5} & Palomar 200-inch & \texttt{DBSP\_DRP}\tnote{6} \\
        KAST\tnote{7} & Shane 3-m & IRAF\tnote{8} \\
        LRIS\tnote{9} & Keck1 & LPipe\tnote{10} \\
        SPRAT\tnote{11} & Liverpool Telescope & \texttt{PypeIt} \\
        NIRES\tnote{12} & Keck2 & \cite{nires} \\
        ESI\tnote{13} & Keck2 & \texttt{makee}\tnote{14} \\
        \bottomrule
    \end{tabularx}
    \begin{tablenotes}
    \item[1] Spectral Energy Distribution Machine \citep{sedm2018}.
    \item[2] \citet{pysedm2019,Kim2022}
    \item[3] Andalucia Faint Object Spectrograph and Camera
    \item[4] \url{https://github.com/jkrogager/PyNOT}
    \item[5] Double Beam Spectrograph \citep{Oke1982}
    \item[6] \texttt{pypeit} \citep{pypeit:joss_pub} based pipeline (\url{https://github.com/finagle29/dbsp_drp}) 
    \item[7] Kast Double Spectrograph \citep{kast}
    \item[8] \citet{Tody1986, Tody1993}
    \item[9] Low Resolution Imaging Spectrometer \citep{Oke1995}
    \item[10] IDL based automatic reduction pipeline (\citealp{Perley2019}; \url{https://sites.astro.caltech.edu/~dperley/programs/lpipe.html)} 
    \item[11] Spectrograph for the Rapid Acquisition of Transients \citep{SPRAT}
    \item[12] Near-Infrared Echellette Spectrometer \citep{nires}
    \item[13] Echellette Spectrograph and Imager \citep{esi2002}
    \item[14] \url{https://www2.keck.hawaii.edu/inst/esi/makee.html}
    
    \end{tablenotes}
    \end{threeparttable}
    \label{tab:inst}
    \hrule
\end{table}

\begin{table}
    \centering
    \caption{Summary of optical and NIR spectra}
    \footnotesize
    \begin{tabularx}{0.41\textwidth}{c c c c}
        \toprule
        \toprule
\textbf{MJD} & \textbf{Phase} & \textbf{Telescope/Instrument} & \textbf{Int} \\ 
& (day) & &(sec)  \\
        \midrule
59972  &  34  &  P60/SEDM  &  2250  \\
60044  &  105  &  LT/SPRAT  &  2200  \\
60056  &  117  &  P60/SEDM  &  1800  \\
60057  &  118  &  P60/SEDM  &  1800  \\
60058  &  119  &  NOT/ALFOSC  &  1800  \\
60058  &  119  &  P60/SEDM  &  1800  \\
60058  &  119  &  Lick-3m/KAST  &  1200  \\
60062  &  123  &  P60/SEDM  &  1800  \\
60063  &  124  &  Keck2/NIRES  &  520  \\
60065  &  126  &  LT/SPRAT  &  2200  \\
60071  &  132  &  P60/SEDM  &  1800  \\
60073  &  134  &  NOT/ALFOSC  &  600  \\
60075  &  136  &  P60/SEDM  &  1800  \\
60081  &  141  &  P60/SEDM  &  1800  \\
60085  &  146  &  Keck1/LRIS  &  300  \\
60087  &  147  &  P60/SEDM  &  1800  \\
60103  &  163  &  P60/SEDM  &  1800  \\
60108  &  168  &  P60/SEDM  &  1800  \\
60110  &  170  &  Keck1/LRIS  &  180  \\
60110  &  170  &  Keck1/LRIS  &  1200  \\
60116  &  176  &  P60/SEDM  &  1800  \\
60116  &  176  &  P200/DBSP  &  600  \\
60120  &  179  &  NOT/ALFOSC  &  1800  \\
60120  &  180  &  P60/SEDM  &  1800  \\
60123  &  183  &  P60/SEDM  &  2250  \\
60124  &  183  &  P60/SEDM  &  1800  \\
60126  &  185  &  P60/SEDM  &  1800  \\
60130  &  189  &  P60/SEDM  &  1800  \\
60138  &  197  &  P60/SEDM  &  1800  \\
60143  &  202  &  Keck1/LRIS  &  300  \\
60144  &  203  &  NOT/ALFOSC  &  2400  \\
60146  &  205  &  P60/SEDM  &  2250  \\
60147  &  206  &  P60/SEDM  &  2250  \\
60148  &  207  &  Keck1/LRIS  &  600  \\
60155  &  214  &  P60/SEDM  &  2250  \\
60161  &  219  &  NOT/ALFOSC  &  2400  \\
60168  &  226  &  P60/SEDM  &  2250  \\
60175  &  233  &  NOT/ALFOSC  &  4400  \\
60199  &  256  &  NOT/ALFOSC  &  1100  \\
60206  &  263  &  Keck2/ESI  &  2700  \\
60224  &  281  &  Keck1/LRIS  &  900  \\
        \bottomrule
    \end{tabularx}
    \label{tab:speclist}
\end{table}

\section{Analysis} \label{sec:analysis}
\subsection{Light curve}

SN~2023aew had a rise of about 2.5 mag over the first 25 days ($\sim$11\,\mphd) to a first peak in TESS data, with a peak magnitude of 17.88~mag in TESS-Red band ($-17.2$; see Figure~\ref{fig:lc}). It then proceeded to decline with an initial decline rate of 2.6\,\mphd in the $r$ band for the next 30 days (rest-frame) and settled onto a slowly declining plateau with only 0.2\,\mphd in the $r$ band between day $60$ to $105$, reaching a minimum brightness of 18.93~mag ($-16.5$). After day $105$, SN~2023aew suddenly started brightening again and rose at a rate of $13.5$~\mphd in the $r$ band and reached a second peak brightness of 16.45~mag ($-18.8$) at day $132$ after which it started turning over to decline at a rate of $\sim$4\,\mphd ($r$ band). Around day $\sim$205, the SN developed a smaller bump in the light curve for $\sim$25 days, before coming back to its previous decline rate. Our last detection of 21.2 mag in the $i$ band was obtained on day 315, and the last limit of $>22.1$ mag in the $r$ band was obtained on day 323. Figure~\ref{fig:lc} also shows the light curves of some peculiar SESNe from the literature for comparison. The light curves of comparison SNe were obtained from the Open Supernova Catalog \citep{opensncatalog2017} for SNe 1993J \citep{P1_1993J,P2_1993J,P3_1993J,P4_1993J,P5_1993J}, 1998bw \citep{P1_1998bw,P2_1998bw,P3_1998bw,P4_1998bw}, 2005bf \citep{tominaga2005,stritzinger2018}, 2007gr \citep{P1_2007gr,P2_2007gr,P3_2007gr}, 2009ip \citep{Mauerhan2013,Margutti2014}, 2011dh \citep{Arcavi2011,Ergon2015} and iPTF13bvn \citep{Fremling2016,P1_13bvn,P2_13bvn}, and from ATLAS forced photometry service \citep{Tonry2018, atlassmith2020,Shingles2021} for SN~2017ens \citep{P1_2017ens}.

\begin{figure}[htbp]
    \centering
    \includegraphics[width=\columnwidth]{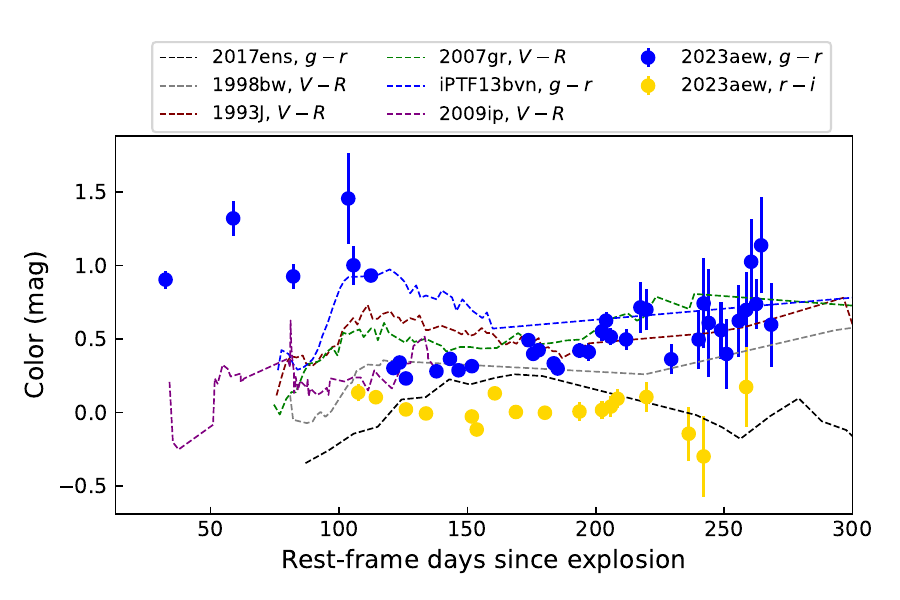}
    \caption{$g-r$ and $r-i$ color curves of SN~2023aew. Shown for comparison are $V-R$ color curves of SNe~1993J, 1998bw, 2007gr, and 2009ip, and $g-r$ color curves of SN~2017ens and iPTF13bvn.}
    \label{fig:color}
\end{figure}

Figure~\ref{fig:color} depicts the $g-r$ and $r-i$ color evolution of SN~2023aew and the comparison SNe. During the declining phase of the first peak, the color of SN~2023aew is red and constant. During the rapid rebrightening the color gets rapidly bluer, then slowly turns red again, similar to the comparison SNe. 

The UV colors obtained with \textit{Swift} just after the second peak (see Figure~\ref{fig:lc}) do not seem particularly bluer than other SESNe at those epochs. However, as there are only a few SESNe observed in UV, and they show a wide variety of behavior in their UV colors, no inferences can be made with certainty \citep{Brown2009,Brown2015}.

\subsection{Bolometric luminosity}\label{sec:bololum}

\begin{figure}[htbp]
    \centering
    \includegraphics[width=0.45\textwidth]{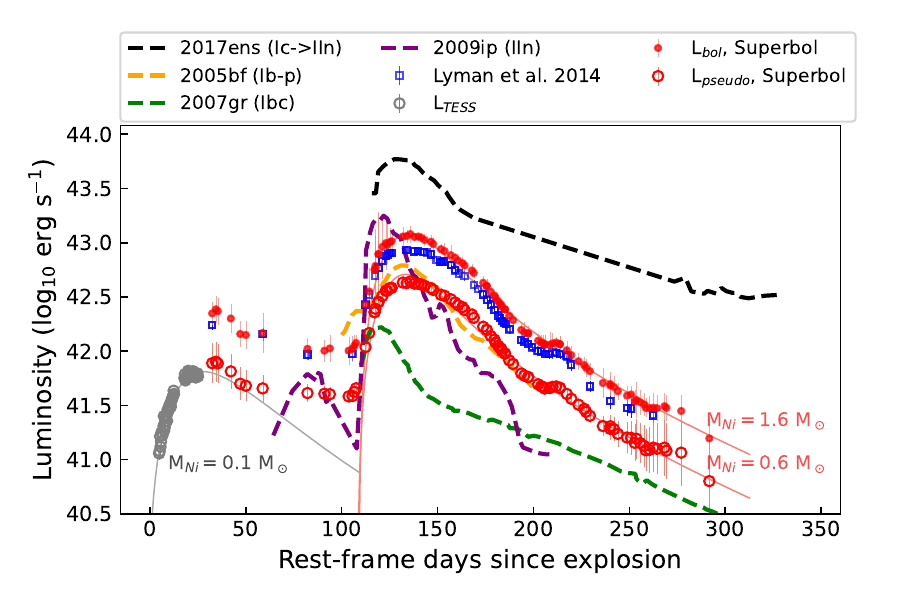}
    \caption{Bolometric (filled red circles) and pseudo-bolometric (open red circles) light curves of SN~2023aew along with some comparison SNe. Gray open circles denote the pseudo-bolometric luminosity approximation from TESS observations. Also shown (blue squares), is an alternative estimate of the bolometric light curve obtained by applying bolometric corrections to $g$-band data following \citet{Lyman2014}.  The $^{56}$Ni radioactivity power fits to the bolometric and pseudo-bolometric light curves following the Arnett method are plotted in light-red color for the second peak and gray color for the first peak.}
    \label{fig:bololc}
\end{figure}

SN~2023aew has good coverage in only TESS-Red band for the first 30 days and then only in $r$ band for the rest of the first peak duration, after which there is decent coverage in all ZTF and ATLAS optical bands. However, as there is no coverage in the UV or the infrared, it is difficult to produce a full bolometric light curve. 

The TESS-Red band fluxes were converted into luminosities by multiplying them with the effective filter width (integrated area under the TESS-Red filter transmission curve; \citealp{svo2012,svo2020}) and the luminosity distance factor.
This estimate was used as an approximate pseudo-bolometric luminosity.

We used \texttt{Superbol} \citep{mnicholl_2018_2155821} with ZTF $gri$ bands and ATLAS $co$ bands to get pseudo-bolometric and bolometric light curves. \texttt{Superbol} interpolates all bands to the $r$-band epochs, calculates pseudo-bolometric luminosity by integrating the observed fluxes over the available bandpasses, and estimates the bolometric luminosity by adding blackbody corrections (absorbed UV, and NIR) to the pseudo-bolometric light curve. Additionally, \texttt{HAFFET} \citep{YangHaffet} was also used to obtain another bolometric light curve estimate by applying bolometric corrections to $g$-band data following \citet{Lyman2014}.

Figure~\ref{fig:bololc} shows both the bolometric luminosity and the pseudo-bolometric luminosity for SN~2023aew along with luminosities of SNe~1998bw \citep[pseudo-bolometric light curve from their figure~18,][]{P3_1998bw}, 2005bf \citep[bolometric light curve from their figure~8,][]{folatelli2006}, 2007gr \citep[pseudo-bolometric light curve from their figure~6,][]{P3_2007gr}, 2009ip \citep[bolometric light curve from their figure~11,][]{Margutti2014}, and 2017ens \citep[pseudo-bolometric light curve from their figure~1,][]{P1_2017ens}. The pseudo-bolometric luminosity of the first peak reached a maximum of $6.6 \pm 0.2 \times 10^{41}$~erg s$^{-1}$, while the second peak reached a maximum of $4.4 \pm 0.6 \times 10^{42}$~erg s$^{-1}$. 
The maximum bolometric luminosity of the second peak is $1.2 \pm 0.2 \times 10^{43}$~erg s$^{-1}$. 

The pseudo-bolometric light curve from TESS data was integrated to obtain the radiated energy output over its duration, and it came out to be $8.0 \pm 0.6 \times 10^{47}$~erg. The bolometric light curve was integrated for the rest of the first peak (until day 100), which came out to be $8.8 \pm 0.4 \times 10^{48}$\,erg. Hence a lower limit of $9.6 \pm 0.5 \times 10^{48}$\,erg can be placed on the total radiated energy during first peak. For the second peak (from rebrightening until our last photometry point at 294 days since explosion), the radiated energy is $5.60 \pm 0.13 \times 10^{49}$\,erg. Thus the total energy radiated by SN~2023aew from explosion until our last detection is $6.56 \pm 0.18 \times 10^{49}$\,erg. Comparison of this radiated energy with other similar events is further discussed in \S\ref{sec:disc:precursor}.

Assuming the two peaks are separate SESNe, we fit $^{56}$Ni power luminosity models using the Arnett \citep{arnett1982,valenti2008} method separately to both peaks. For the first peak, we use the pseudo-bolometric luminosity from TESS observations and from Superbol (see above) covering day 0 to 50 since explosion to fit the Arnett radioactivity model. We obtain a lower limit on nickel mass, $M_{\mathrm{Ni}}=0.11^{+0.02}_{-0.06}$\,\Msun\ and ejecta mass, $M_{\mathrm{ej}}=27.6^{4.1}_{-19.0}$\,\Msun\ assuming a photospheric velocity of 11,800~\kms (see \S\ref{sec:firstpeak}).

Next, assuming that the second brightening is also powered by $^{56}$Ni decay, $gri$ data during the rise of the second peak were fitted with power laws using \texttt{HAFFET} to obtain an ``explosion" epoch, which came out to be $\sim$115 days after the explosion epoch from TESS. Using this explosion epoch as a reference, models were fitted to the bolometric and pseudo-bolometric light curves respectively, which seem to agree well except around the bump at 210 days from explosion. The radioactivity power fit to $L_{\mathrm{bol}}$ requires a nickel mass, $M_{\mathrm{Ni}}=1.59^{+0.62}_{-0.40}$\,\Msun\ and an ejecta mass, $M_{\mathrm{ej}}=8.52 \pm 2.40$\,\Msun\ assuming a photospheric velocity of 6,000~\kms (see Figure~\ref{fig:HeHO}). A fit to the pseudo-bolometric luminosity provides a lower limit of $M_{\mathrm{Ni}}=0.59^{+0.31}_{-0.19}$\,\Msun\ and $M_{\mathrm{ej}}=7.62 \pm 3.16$\,\Msun. Clearly, this nickel mass estimate is unreasonably high compared to 
what is observed in other SESNe and what is predicted from models, and thus must be hinting at an additional power source for the second peak.

Also, the radioactivity models for the ``two SESNe" combined cannot explain the luminosity of the plateau that bridges the two peaks (see Figure~\ref{fig:bololc}), thus making the two separate SESNe scenario less likely.


\subsection{First peak of SN~2023aew}\label{sec:firstpeak}
\begin{figure}[htbp]
    \centering
    \includegraphics[width=\columnwidth]{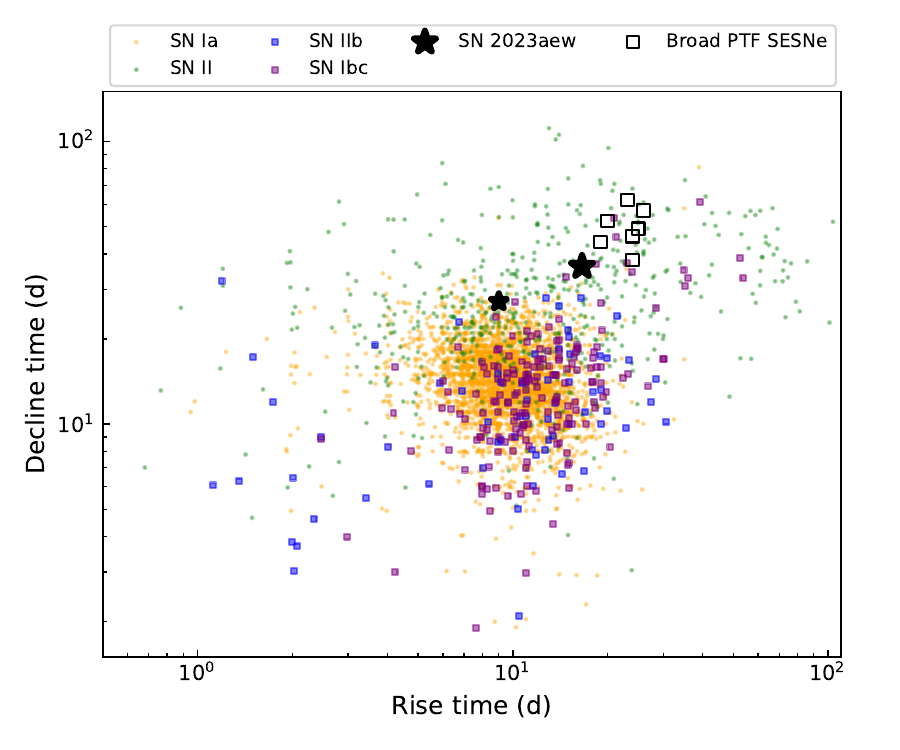}
    \caption{Rise time versus decline time (peak to half-peak in the $r$ band) of the Bright Transient Survey sample of supernovae (colored points) and broad light-curve SESNe from \citet{Emir2023} (empty black squares). The first peak of SN~2023aew is marked with the smaller black star (rise time from TESS data) and the second peak is marked with the larger black star ($r$ band data).}
    \label{fig:risefall}
\end{figure}

Serendipitous coverage from TESS revealed the explosion epoch, the rise and the peak of the first bump which were not detected in any other data. The overall light curve shape of SN~2023aew resembles SN~2009ip but is much broader in both peaks (see Figure~\ref{fig:lc}). Figure~\ref{fig:lc} also compares the first peak with Type IIb SNe~1993J and 2011dh, which have a faster decline than SN~2023aew and narrower light curves. The rise time of the first peak from explosion epoch to peak is 20 rest-frame days and from half-peak flux to peak flux is $\sim$9 rest-frame days. The decline time from peak to half-peak is $\sim$27 rest-frame days but is likely affected by the plateau that the first peak develops at 50 days. The half-peak to peak rise and decline times of the first peak are compared with a sample of bright supernovae obtained from the ZTF Sample Explorer\footnote{\url{https://sites.astro.caltech.edu/ztf/bts/explorer.php}} (\citealp{Perley2020,Fremling2020}; classified since the start of ZTF and having pre- and post-peak coverage) in Figure~\ref{fig:risefall}. The first peak rise seems consistent with the BTS sample SNe, but the decline time is slightly higher than for the SESN sample and more towards the SN~II population.

\begin{figure*}[h]
    \centering
    \includegraphics[width=0.98\textwidth]{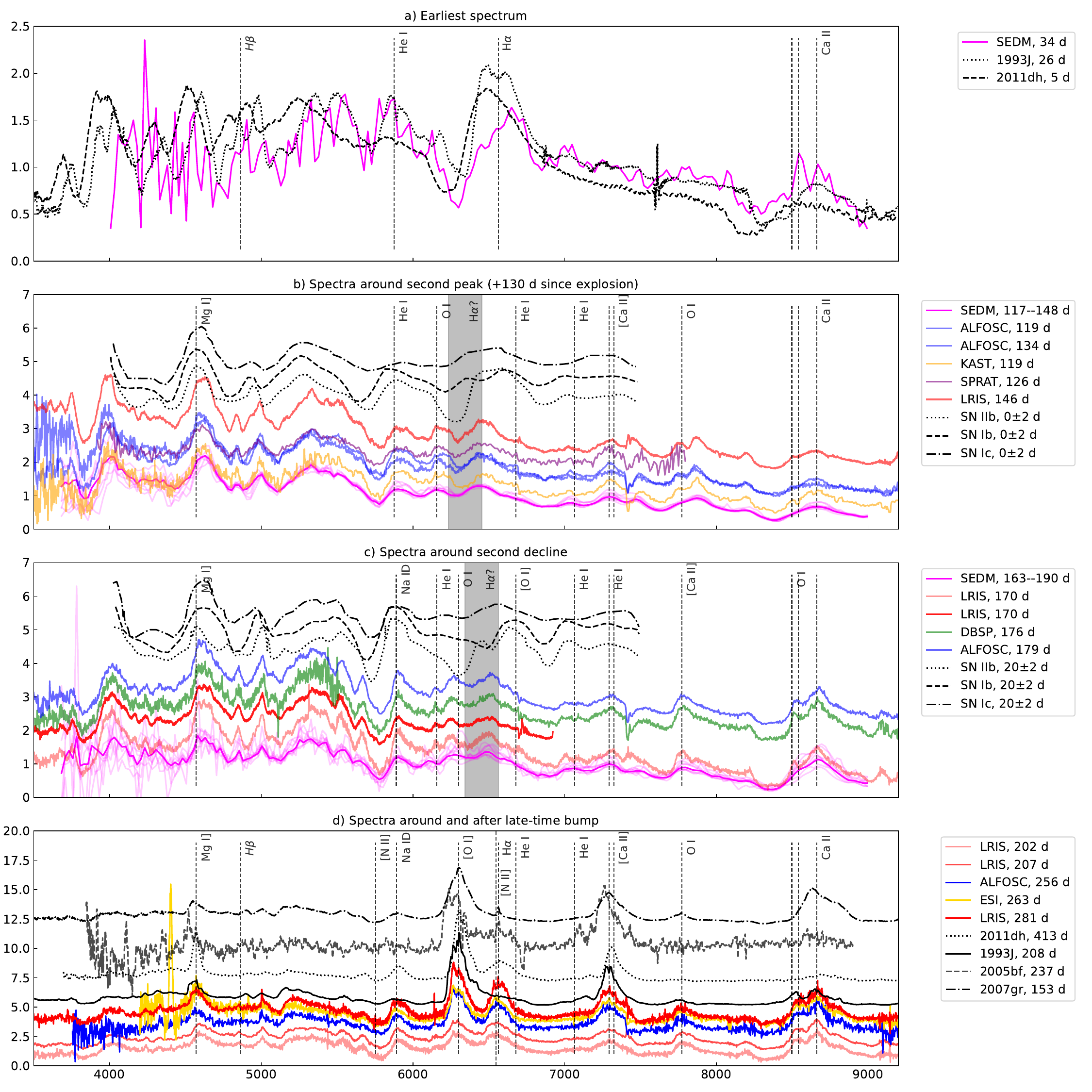}
    \caption{SN~2023aew spectral comparison at different phases compared to mean spectra of SNe IIb, SNe Ib and SNe Ic from \citet{liu2016}.}
    \label{fig:speccomp}
\end{figure*}

After ZTF discovery, an initial spectrum at the first peak was obtained through the usual spectroscopic efforts of the Bright Transient Survey 
which showed a P-Cygni H$\alpha$ profile of with a velocity of $\sim$11,800 \kms. Using the Python version of template matching supernova classification software \texttt{Superfit} \citep{superfit,ngsf}, a match to the Type IIb SN~2001ig was obtained and SN~2023aew was classified as a SN IIb (\S\ref{sec:intro}). Panel (a) of Figure~\ref{fig:speccomp} shows that this earliest spectrum of SN~2023aew ($+34$ days from explosion) is a good match with the early time spectra of Type IIb SNe~1993J ($+28$ days) and 2011dh ($+5$ days). The broad component of H$\alpha$ is similar in velocity with the broad component of the precursor spectra of Type IIn SN~2009ip, however any narrow components if present cannot be discerned in our SEDM spectrum. The deviation from 09ip-like behavior occurs during the second peak, where SN~2023aew transforms into a SN~Ibc and does not show any narrow lines in the spectra.

\subsection{Second peak of SN~2023aew}

 For the second peak resulting from the rapid rebrightening, the rise time from half of the peak luminosity to peak luminosity is $\sim$17.5 rest-frame days, and the decline time from the peak to half of the peak luminosity is $\sim$36.1 rest-frame days. The light curve width at half-max is $\sim$53.5 rest-frame days. These values are higher than what has typically been observed for normal SESNe \citep{prentice2016,taddia2018}. In Figure~\ref{fig:risefall}, the half-peak to peak rise and decline times of the second peak of SN~2023aew are compared with the BTS sample and with the sample of broad light-curve SESNe from Palomar Transient Factory (PTF) presented in \citet{Emir2023}. The second peak of SN~2023aew is indeed broader than what is the case for most ZTF SESNe and consistent with those in the broad \citet{Emir2023} sample. Though none of the SESNe in \citet{Emir2023} had a long precursor-like first peak as SN~2023aew, some of them do display similar undulations in the light curve. \citet{Emir2023} favor ejecta from massive stars ($>20-25~\Msun$) as the cause behind the broad light curves but do not rule out hidden CSM interaction or additional powering mechanisms.

Figure~\ref{fig:speccomp} panel (b) compares spectra taken around the second peak maximum of SN~2023aew to the mean spectra of SNe IIb, SNe Ib, and SNe Ic at peak ($0\pm2$\,days) as constructed in \citet{liu2016}. The absorption feature at \ion{He}{1} $\lambda5876$ appears to be closer in strength to a SN Ib rather than a SN Ic, with perhaps a weak helium feature present also at 7065\,\AA. On 
the other hand, with \ion{He}{1} $\lambda\lambda6678$ and 7065 being weak or absent, the 5876\,\AA\ feature could be due to \ion{Na}{1}D instead, making a stronger case for a spectral similarity to SNe Ic. The presence of trace helium in SNe Ic is also highly debated \citep{branch2002,elmhamdi2006}, however, \citet{liu2016} suggests that for a true SN Ib classification either the 5876\,\AA\ line should be strongly identified before maximum or all three \ion{He}{1} lines (5876, 6678, 7065 \AA) should be present post maximum and at $<40$ days. Considering the phase of SN~2023aew to be near maximum (and second brightening to be the main peak) and given that the 5876\,\AA\ line is clearly present, we suggest that SN~2023aew resembles more a SN Ib at this phase. 
There is an absorption line around $\sim$6200\,\AA\ that matches the H$\alpha$ absorption from the earliest spectrum, however, the corresponding emission peak is blueshifted from the H$\alpha$ rest wavelength and redshifted from \ion{Si}{2} $\lambda6355$. Several studies \citep{matheson2001,branch2002,elmhamdi2006,yoon2010,liu2016,parrent2016} have indicated the presence of trace hydrogen in SNe Ib and that the origin of similar features around 6000--6400\,\AA\ in SNe Ib could be blueshifted emission of H$\alpha$ \citep[their \S3.1,][]{Gal-Yam2017}. 

Figure~\ref{fig:speccomp} panel (c) compares post-second peak decline spectra of SN~2023aew with \citet{liu2016} SESN mean spectra at $20\pm2$\,days past maximum. The spectral evolution of SN 2023aew is slow, but at this stage it has started developing [\ion{O}{1}] $\lambda$6300 and strong \ion{Ca}{2} NIR emission. Absorption near \ion{He}{1} $\lambda$5876 is strong but likely \ion{Na}{1}D at this stage and the other \ion{He}{1} lines are much weaker compared to the SN Ib mean spectrum. The overall spectra at this stage appear more similar to the SN Ic than to the SN Ib template, which is also supported by SNID matches to SN Ic templates \citep{aew_astronote2}.

\begin{figure}
    \centering
    \includegraphics[width=\columnwidth]{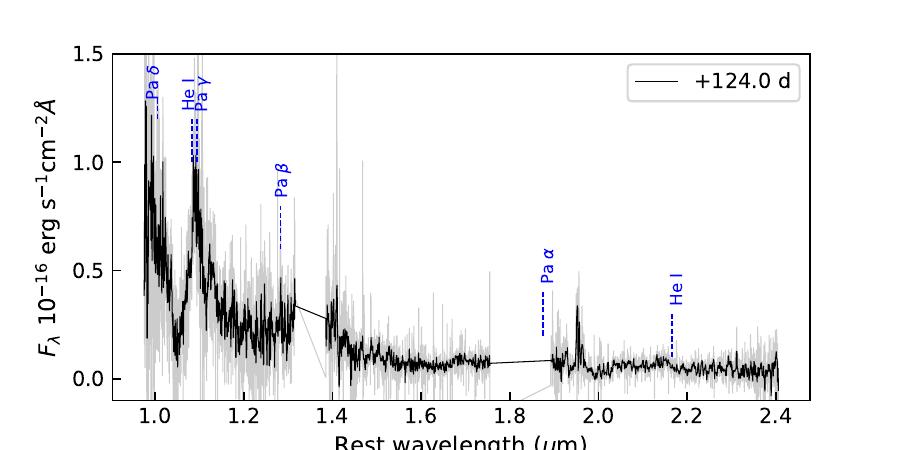}
    \caption{A near-infrared spectrum of SN~2023aew obtained with Keck2/NIRES, smoothened using a median filter of kernel size of 9 pixels (in black). Tentative line identifications are marked with blue dashed lines.}
    \label{fig:nires}
\end{figure}

A near-infrared spectrum was also obtained with Keck2/NIRES covering a wavelength range from 1.0--2.4~$\mu$m but unfortunately the data has poor signal-to-noise ratio. Figure~\ref{fig:nires} shows the NIR spectrum along with some possible line identifications. The Paschen series is marked, as well as \ion{He}{1} around 1.085~$\mu$m. Pa$\alpha$ falls in a no-coverage zone, and Pa$\beta$ and Pa$\delta$ fall in high noise regions, hence are not clearly discernible in the spectrum. The feature near 1.1~$\mu$m could be either Pa$\gamma$ or \ion{He}{1} but the helium line at 2.2~$\mu$m is not detected. 

\subsection{Nebular spectra and emergence of H$\alpha$}

Figure~\ref{fig:speccomp} panel (d) compares spectra taken around the late-time bump at $\sim200$ days from explosion to nebular spectra of SESNe 1993J (IIb), 2011dh (IIb), 2005bf (Ib-pec) and 2007gr (Ic) obtained from the Open SN catalog. The strongest features present at this stage are \ion{Ca}{2} NIR, [\ion{Ca}{2}] $\lambda\lambda$7292, 7324, \ion{O}{1} $\lambda$7774, [\ion{O}{1}] $\lambda\lambda$6300, 6364, \ion{Na}{1}D $\lambda$5890 and an emission feature centered around H$\alpha$. [\ion{N}{2}] $\lambda\lambda6548, 6583$ is also a major contributor of flux around the H$\alpha$ wavelength at the nebular phases in the low mass end of SESNe (low mass Type IIb, \citealp{jerkstrand2015}), but almost absent for the high mass SESNe. The same is reflected in the spectra of comparison SESNe in Figure~\ref{fig:speccomp} panel (d), with SN~2011dh (IIb) having the most flux in the [\ion{N}{2}] line and SN~2007gr (Ic) barely having any. However, SN~2023aew seems to have a larger flux in that line compared to the others, possibly due to contribution from H$\alpha$. H$\beta$ is almost non-existent, but is also similarly weak in the day 208 spectrum of SN~1993J. 

\begin{figure}[htbp]
    \centering
    \includegraphics[width=0.47\textwidth]{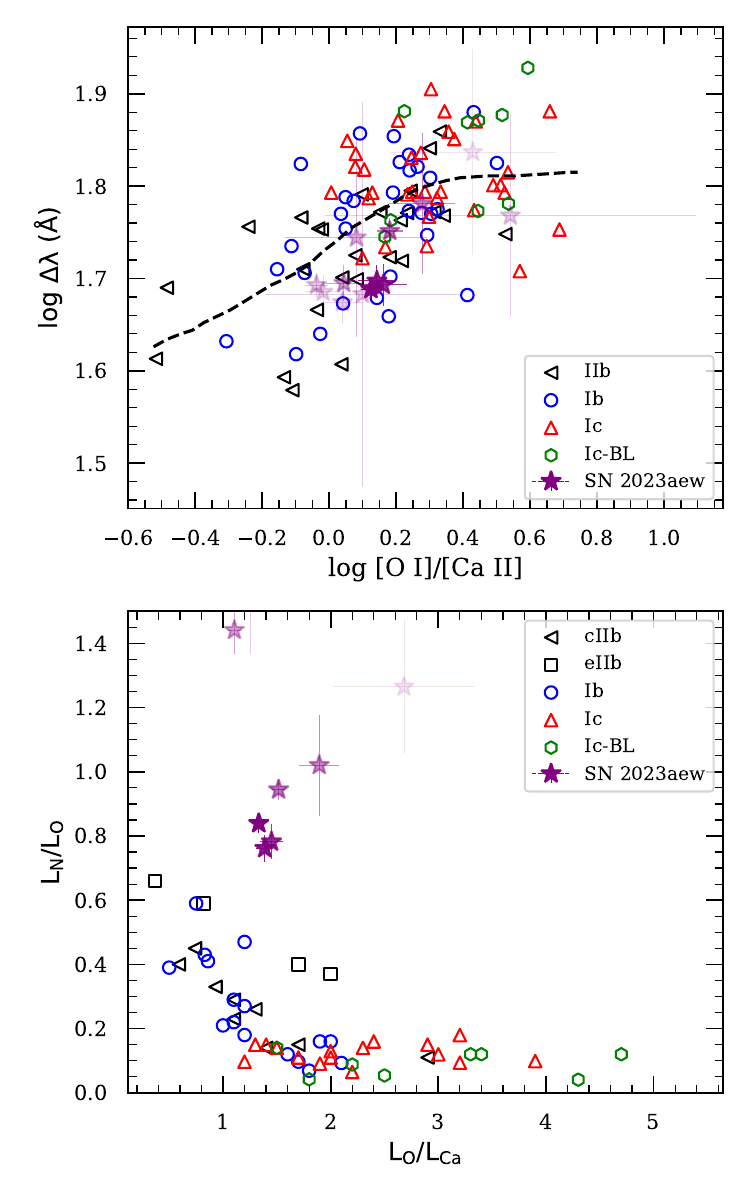}
    \caption{Upper panel: SN~2023aew in the logarithmic phase-space of [\ion{O}{1}] width versus [\ion{O}{1}]/[\ion{Ca}{2}] flux ratio for SESNe published in \citet[their fig.\,7a]{fang2022}. Lower panel: $L_{\mathrm{N}}$/$L_{\mathrm{O}}$ versus $L_{\mathrm{O}}$/$L_{\mathrm{Ca}}$ for SESNe published in \citet[their fig.\,3]{fang2019}. `cIIb' and `eIIb' refer to compact and extended SNe IIb respectively. Values for SN~2023aew are shown in purple stars and derived from late-time spectra (phases ranging from 200 to 281 days from explosion) with the transparency decreasing with increasing phase. The SN nebular spectra used in \citet{fang2019} range from 150 to 300 days after the peak while for SN~2023aew they range from 75 to 150 days after the second peak. SN~2023aew consistently shows higher values on the [\ion{N}{2}]/[\ion{O}{1}] axis, which suggest the presence of H$\alpha$ contaminating (or rather dominating) the [\ion{N}{2}] lines at these phases.}
    \label{fig:neb1}
\end{figure}

To further explore the nature of SN~2023aew, flux ratios of nebular diagnostic lines are compared with the analysis of SESNe presented in \citet{fang2019,fang2022}. The line fluxes of [\ion{O}{1}] $\lambda\lambda6300, 6364$, [\ion{Ca}{2}] $\lambda\lambda7292, 7324$ and [\ion{N}{2}]/H$\alpha$ complex, and the width of the [\ion{O}{1}] line are calculated for SN~2023aew following similar procedures as in \citet{fang2019,fang2022}. Figure~\ref{fig:neb1} top panel plots the correlation of the [\ion{O}{1}]/[\ion{Ca}{2}] ratio versus the [\ion{O}{1}] width for SN~2023aew along with data obtained from figure 7a of \citet{fang2022}. Figure~\ref{fig:neb1} bottom panel plots the line ratio of [\ion{N}{2}]/[\ion{O}{1}] versus the line ratio of [\ion{O}{1}]/[\ion{Ca}{2}] for SN~2023aew along with data obtained from figure 3 of \citet{fang2019}. The [\ion{O}{1}] fluxes were calculated by fitting and subtracting a pseudo-continuum to the oxygen-nitrogen complex, then fitting and subtracting a Gaussian profile centered at 6563\,\AA\ to remove the [\ion{N}{2}]/H$\alpha$ contribution, and finally integrating the remaining flux in the complex over a suitable wavelength range. The [\ion{N}{2}] fluxes were similarly calculated by subtracting a Gaussian profile fit centered at 6300\,\AA\ to remove the [\ion{O}{1}] contribution after continuum removal. The [\ion{Ca}{2}] fluxes were calculated after subtracting a pseudo-continuum and integrating over the line. The uncertainties were calculated using the Monte-Carlo method as described in Appendix~\ref{sec:app}. In both panels, the measurements for SN~2023aew are marked with purple stars having decreasing transparency with increasing phase.

\citet{fang2022} discerned that oxygen-rich ejecta expand faster and the SESN subtype distribution showed that SNe IIb/Ib are more steeply correlated than SNe Ic/Ic-BL (see their figure 7a and top panel of our figure~\ref{fig:neb1}). SN~2023aew has a log$_{10}$([\ion{O}{1}]/[\ion{Ca}{2}])~$\sim$~0.1, before the correlation curve in Figure~\ref{fig:neb1} (top panel) starts to flatten, but has slower velocities, again more towards the phase space that SNe~IIb/Ib occupy. The luminosity ratio of [\ion{N}{2}]/[\ion{O}{1}] in Figure~\ref{fig:neb1} (bottom panel) is considerably higher than the corresponding values \citet{fang2019} measured for SESNe at that L$_O$/L$_{Ca}$, once again suggesting the presence of hydrogen in the nebular phase. 

\begin{figure*}[ht]
    \centering
    \includegraphics[width=0.95\textwidth]{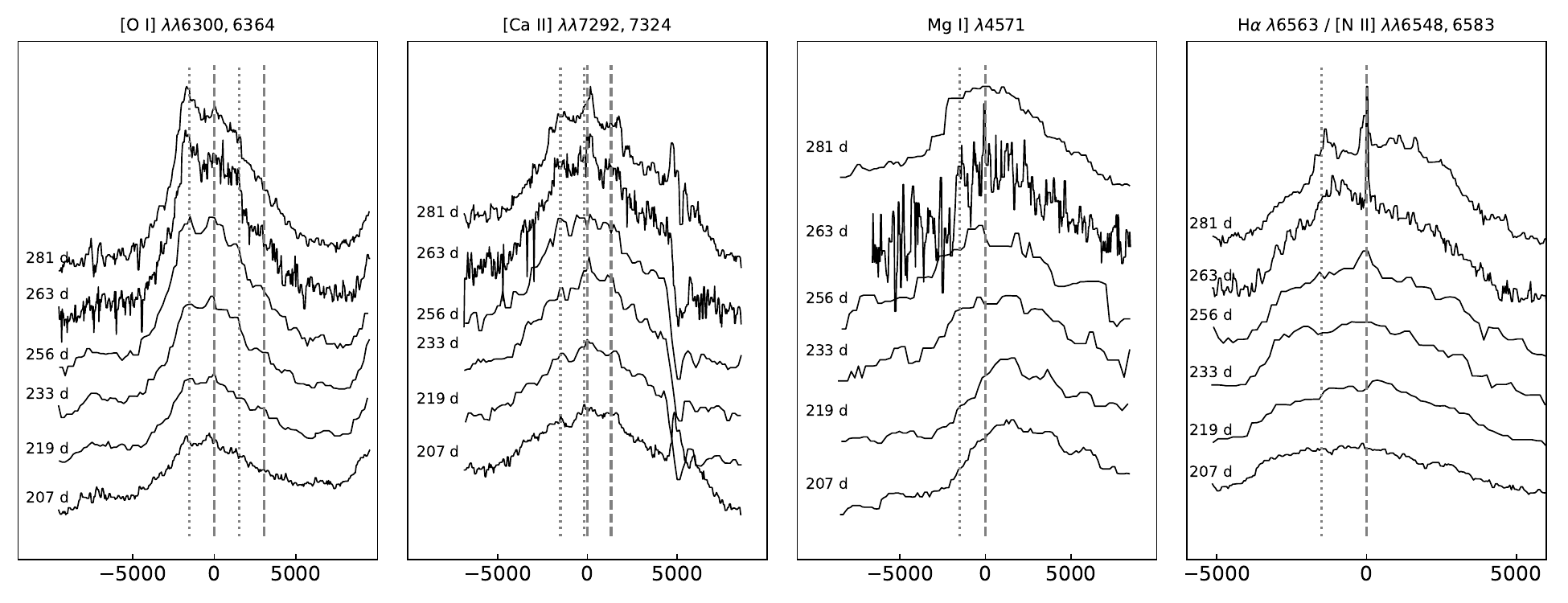}
    \caption{Line profiles of nebular [\ion{O}{1}], [\ion{Ca}{2}], \ion{Mg}{1}] and H$\alpha$/[\ion{N}{2}] complex from 207 days to 281 days. Dashed vertical lines mark the rest wavelengths and dotted vertical lines mark the blueshifted ($\sim$1500 \kms) components. The blueshifted component in all lines appears to be increasing in strength with phase.}
    \label{fig:lineprofile}
\end{figure*}

Further arguments for the 6500 Å feature consisting mostly of hydrogen are found when comparing to Barmentloo et al. (2024, in preparation). In this work, the authors compare [\ion{N}{2}] emission from a set of nebular model spectra to a sample of observed SESNe. They find that up until $\sim$ 200 -- 250 days post-explosion, the contribution of [\ion{N}{2}] in this region is below 50$\%$ for all progenitor models, with the feature being almost non-existent for $M_{\text{preSN}} \gtrsim 4.5$ \Msun{}. Perhaps more importantly, it is found that the line widths for the [\ion{N}{2}] feature in their sample have a lower limit FWHM of 170 Å, with the median around $\sim$ 200 -- 220 Å. Performing the same analysis for SN 2023aew, we find FWHM values between 115 -- 135 Å. This means that if the feature would be mostly [\ion{N}{2}], SN 2023aew would be a significant outlier (as when comparing with \citealt{fang2022}). Finally, calculating the $f_{[N II]}$ diagnostic from Barmentloo et al. (2024) for our spectral series, the resulting values would indicate a $M_{\text{preSN}} \lesssim$ 3 \Msun{}, which does not match with any of our earlier estimates. All in all, the above findings suggest that any presence of [\ion{N}{2}] could only be very minor, with the majority of the emission in this region due to other elements, most probably hydrogen.

A final inference that can be derived from the nebular spectra comes from the line profiles of [\ion{O}{1}] $\lambda\lambda6300, 6364$, [\ion{Ca}{2}] $\lambda\lambda7292, 7324$ and \ion{Mg}{1}] $\lambda4571$. Figure~\ref{fig:lineprofile} shows the nebular line profiles with dashed vertical lines marking the rest line wavelengths and dotted vertical lines marking blueshifted (by 1500\,\kms) wavelengths. The [\ion{O}{1}] profile clearly has blueshifted components for both the 6300 and 6364\,\AA\ lines creating a double-peaked profile. [\ion{Ca}{2}] also has hints of double-peaked structure with peaks blueshifted by 1500\,\kms, though not as clear as for [\ion{O}{1}]. The \ion{Mg}{1}] peak is redshifted at earlier epochs but a blueshifted component seems to develop at later times. The blueshifted peak in all three lines seems to get stronger with time. The double-peaked shape can also be discerned in \ion{O}{1} $\lambda7774$ and in the H$\alpha$/[\ion{N}{2}] complex for some epochs. Double-peaked structure in [\ion{O}{1}] lines has been observed in nebular spectra of many SESNe \citep{modjaz2006, modjaz2007,maeda2007,modjaz2008}, and in most cases, the two peaks are symmetric around the rest wavelength (for example see \citealp[their figure 2]{modjaz2008}). SN~2005bf \citep{folatelli2006,maeda2007}, a peculiar Type Ib SN was an exception with a highly blueshifted ($\sim$2000\,\kms) trough, similar to what is seen here for SN~2023aew. An aspherical explosion in the shape of a torus could give rise to a double-peaked feature in the case of optically thin ejecta at nebular times, but 
does not
explain the blueshift. For SN~2005bf, a unipolar blob of low mass, accelerated by a pulsar kick was suggested by \citet{maeda2007}. On the other hand, less extreme blueshifts can be explained by residual opacity effects as described in \citet{taubenberger2009}. More recently, \citet{fang2023} analyzed a sample of nebular spectra of SESNe and found that roughly half of the SESNe had either a double-peaked [\ion{O}{1}] or [\ion{Ca}{2}] but none had double-peaked structure in both lines. \citet{fang2023} theorize an axisymmetric model for their observations, where the oxygen-burning ash (Ca-rich region) is distributed in bipolar bubbles with unburnt oxygen outside it and depending on the viewing angle leads to a double-peaked profile in one of the lines. However, applying this prescription to SN~2023aew is difficult due to the structure present in the calcium line and the blueshifted trough. More advanced 3D models are just beginning to be explored \cite[e.g.,][]{vBaal2023}.

\section{Discussion} \label{sec:discussion}
\subsection{Two distinct supernovae or a single peculiar one?}

To determine whether the light curve of SN~2023aew is due to a single transient or potentially due to two separate events, like two SNe that just happened to explode very close in time and space (along the line-of-sight), we imaged SN~2023aew with WaSP (Wafer-Scale Imager for Prime) on the Palomar 200-inch telescope (P200) in $g$, $r$ and $i$ bands 144 days after explosion, right around the second peak. The exposure time used per band was 300 seconds, which corresponds to a 5$\sigma$ limiting magnitude of $\sim22.5$ mag for WaSP in good seeing conditions. The SN was observed at an average seeing of 1.4$^{\prime\prime}$. On the same night, a confirmed sibling SN~Ia pair, SN~2023egs, with peaks separated by $\sim$20 days and on-sky separation of $\sim1.6^{\prime\prime}$ was also observed with WaSP. The first SN Ia of this pair was $\sim$50 days past maximum at the time of observation and thus would be at least 3 magnitudes fainter than peak (18.4~mag) if it was a normal SN~Ia \citep{Phillips2017}, making it $\gtrsim$ 21~mag. The second SN was $\sim$19~mag at the time of observation and both SNe~Ia in SN~2023egs were clearly detected in the WaSP image. Considering the initial decline of the first peak of SN~2023aew (2.6 \mphd), the ``first" SN in SN~2023aew at 144 days from explosion would be $\sim21.1$ mag, while the second SN was $\sim$16.7~mag and thus both would still be separately detected in WaSP data if they were $\sim2^{\prime\prime}$ apart.    
Upon analyzing the WaSP observations of SN~2023aew, there did not seem to be two sources present at the location of SN~2023aew within the seeing limit. 

\citet{Melissa2022} analyzed sibling SNe (SNe sharing the same host galaxy) in the ZTF Bright Transient Survey sample of 2 years and found 5 sibling pairs (10 SNe) brighter than 18.5 mag, corresponding to a rate of only $\sim1\%$ of total SNe in BTS in that same period. Out of these, only 2 were SESNe, i.e. about one per year. The lowest on-sky separation of these siblings was 3.7$^{\prime\prime}$. Thus, the chances of SN~2023aew being a SESN sibling pair with coincident location ($<2^{\prime\prime}$) and also exploding within $\sim115$ days (4 months) of each other 
is extremely small. 

Post-facto statistics is difficult, but
another crude estimate of the rarity of SN~2023aew being a sibling pair might be calculated as follows. Assuming a SN rate of 1 per 100 years per galaxy, the Poisson probability of two SNe exploding within half a year is $P_1\sim 1\times 10^{-5}$. As SN~2023aew is roughly 7~kpc from the center of its host, the probability of the siblings occurring at that radius also needs to be taken into account. This probability can be estimated by taking the radial distribution of CCSNe in a galaxy from \citet[their fig.\ 2]{Wang1997}, then calculating the ratio of the integrated distribution over a radius span of 6--8 kpc with integrated distribution over the galaxy span (assuming $\sim20$ kpc), which comes out to be $P_2\sim0.14$. From the P200/WaSP image, the on-sky separation is known to be within 2$^{\prime\prime}$, which translates to $\sim1$ kpc at the distance of SN~2023aew. Thus the fraction of volume of the 6--8 kpc disc that the siblings are expected in would be at most $P_3\sim0.01$. 
The exact off-centre distance to the host is of course not important, but the fact that the SN exploded in the outskirts of a resolved galaxy where the star formation rate is limited makes the probability for two unrelated SNe much more unlikely.
Thus the total probability of two SESNe exploding within 0.5 year in the same galaxy at $\sim7$~kpc from the galaxy center and within 2$^{\prime\prime}$ on-sky separation is $P\sim1.7\times10^{-8}$ per galaxy. Next, the maximum distance out to which SN~2023aew ($M_{\mathrm{abs}}^{\mathrm{peak}}\sim-18.7$) would be detected and classified by the BTS survey (flux limit of 18.5 mag) is $\sim275$ Mpc. Taking the density of galaxies in this volume (assuming MW-like) to be $\sim0.006\ \mathrm{Mpc}^{-3}$, assuming a uniform distribution and accounting for the fact that ZTF can only observe $\sim0.75$ of the sky, the number of galaxies ZTF will observe in 275 Mpc volume is $\sim392000$. Multiplying that with the per galaxy probability, the expected number of siblings like SN~2023aew is $\sim0.007$ and hence the Poisson probability of detecting one event is $0.7\%$. 
Even more unlikely is probably that both of these events are SESNe, and in particular the unusual properties of the second peak in terms of lower-than-average line velocities and broader-than-average and more luminous peak makes it very unlikely that the events are not linked. We also do not know of any mechanism that would make one SN trigger the other in a common system on such short timescales.

\subsection{Rebrightening or precursor?}\label{sec:disc:precursor}

The usual interpretation for supernova rebrightening is that it is caused by interaction with a CSM shell ejected during the final moments of the progenitor. This is most frequently observed in SNe IIn, where the light curves have multiple undulations and late-time emission, see for example \citet{Nyholm2017}. SN~2021qqp \citep{Hiramatsu2023} had a long, slow-rising precursor before the first peak as well as a late-time brightening after about a year, with both first and second peaks showing spectral similarity. Late-time emission due to CSM interaction has also been observed in spectrally normal SESNe that develop narrow emission lines and secondary light-curve plateaus or peaks, some examples being SNe~2017ens \citep{P1_2017ens} and 2019oys \citep{sollerman2020}. In the case of SN~2023aew, the second peak is spectrally different (Ibc-like) from the first (IIb-like) unlike SN~2021qqp and even though the SN light curve shows undulations, it does not evolve into having interaction-dominated spectra with narrow lines like SNe~2017ens and 2019oys. Hydrogen is present during the first peak but not during the second one and then comes back again in the nebular phase but has broader velocities than strongly interacting SNe IIn. If CSM interaction is indeed the cause of the rebrightening, the corresponding spectral signs are hidden. The horned and blueshifted [\ion{O}{1}] and [\ion{Ca}{2}] emission line profiles (see Figure~\ref{fig:lineprofile}) which could arise due to asymmetric gas distribution (SN~2005bf; \citealp{tominaga2005,maeda2007,modjaz2008}) also hint at unusual geometry that could possibly hide the spectral signatures of interaction. \citet{sollerman2020} presented two interacting SESNe, one where the dramatic light curve transformation was accompanied with spectral interaction signatures while the second only depicted slight undulations. Hence it is less likely for such a dramatic rebrightening to not be associated with transformation into interacting SN.

Other suggested causes of double peaks include double-peaked nickel distribution in the ejecta (SN~2005bf; \citealp{tominaga2005}) and delayed magnetar energy injection \citep{maeda2007}. However the timescales from both of these scenarios are not consistent with the evolution of SN~2023aew \citep{Kasen2016,Orellana2022}. 

\begin{figure}[htbp]
    \centering
    \includegraphics[width=0.47\textwidth]{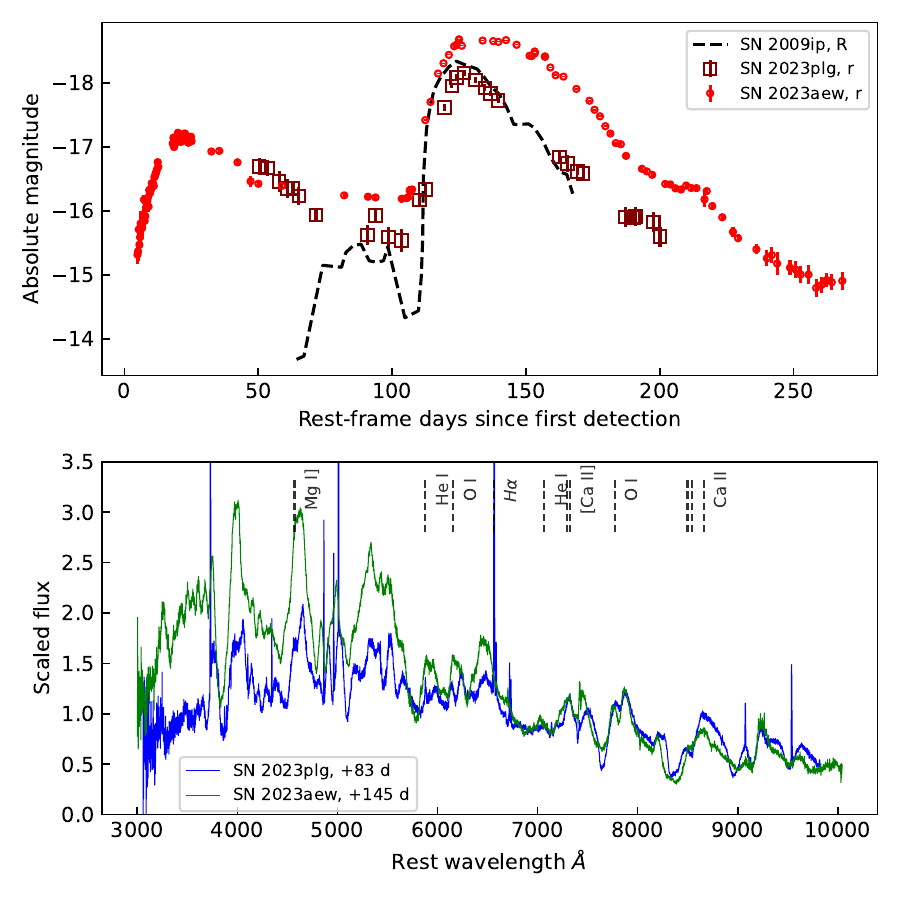}
    \caption{Comparison of SN~2023aew with the canonical precursor Type IIn SN 2009ip, but also with the analogous SESN SN~2023plg showing both a similar luminous, long-lived declining precursor and a bright second peak. Light curves shifted to match the rise of the main peak. SN 2023plg also exhibits a Type Ibc SN spectrum.}
    \label{fig:xuvkn}
\end{figure}

Another possibility is the first peak being a precursor emission \citep{Ofek2014,Strotjohann2021} similar to SN~2009ip \citep{Mauerhan2013,Pastorello2013,Margutti2014}. Pre-cursors have also been seen in hydrogen-deficit SNe \citep{Brennan2024}. Figure~\ref{fig:xuvkn} upper panel compares the light curves of SN~2023aew (circles) and SN~2009ip (dashed lines) and many similarities are apparent. While SN~2023aew has a more luminous and longer precursor than SN~2009ip, the following main peak is also more luminous and broader, and both SNe show undulations in their declining light curves. The main peak rise time is also similar in both SNe. However, the SN~2023aew precursor is much redder than the SN~2009ip precursor (see Figure~\ref{fig:color}). SN~2009ip also showed clear narrow, intermediate, and broad velocity features in the H$\alpha$ emission line and thus it was evident that circumstellar material was present and interacting with the SN ejecta, it was a clear Type IIn SN. However, for SN~2023aew, given that the only spectrum taken during the precursor event is from P60/SEDM, only the broad velocity feature can be resolved (11,800 \kms), which is similar to the broad velocity seen in SN~2009ip. 

If we assume for a moment that SN~2023aew is similar to a 09ip-like event, with the first peak being due to an eruption (a precursor) and the second peak the actual supernova explosion. The longer and much more energetic precursor outburst could be generating energy when the SN ejecta later run into this material to power the second peak in addition to radioactive $^{56}$Ni. An eruption mechanism could be considered for the precursor following the models in \citet{MM2022}, wherein the precursor light curve is recombination-driven similar to Type IIP SNe. Though not observed in the first peak of SN~2023aew, perhaps due to inadequate spectral resolution, we assume a similar narrow line velocity as for SN~2009ip ($\sim$1,500 \kms) in the following calculations. 

\citet{MM2022} found that their semi-analytical models largely follow the \citet{Popov1993} analytical scalings \citep[eqs.\ 19,20][]{MM2022}. For SN~2023aew, the ``precursor" plateau duration ($t_{\mathrm{pl}}$) is 80 days (which is a lower limit considering possible interruption by the following SN explosion), and the precursor radiated energy is $E_{\mathrm{pl}} \approx 1.0 \times 10^{49}$ erg which gives a precursor plateau luminosity ($L_{\mathrm{pl}} = \frac{E_{\mathrm{pl}}}{t_{\mathrm{pl}}}$) of $1.5 \times 10^{42}$ erg\,s$^{-1}$. Then using the inverted Popov equations and an ejecta speed of $v_{\mathrm{ej}} \approx 1500$ \kms, we estimate an ejecta mass of $M_{\mathrm{ej}} \approx 0.61$ \Msun\ and an initial radius of $R_0 \approx 9774$ \Rsun\ for the precursor, both values are comparable to the SNe analyzed in \citet{MM2022}. Then again from \citet[eqs.\ 27,28]{MM2022}, we obtain an outer radius of the precursor ejecta 
$R_{\mathrm{CSM}} \sim 1 \times 10^{15}$ cm, 
and density of the precursor ejecta $\rho_{\mathrm{CSM}} \sim 2.6 \times 10^{-13}$ g\,cm$^{-3}$, which is a more extended and less dense CSM than for SN~2009ip \citep[see fig.\ 8,][]{MM2022}. Such a CSM, coupled with asymmetric geometry (signatures of which are present in the form of double-peaked nebular lines) could potentially result in a SN-CSM interaction without narrow lines in the optical spectra. Estimating the luminosity from shock heating using \citet[][their eq.\ 29]{MM2022}, we get $L_{\mathrm{sh}} \approx 2 \times 10^{43}$ erg\,s$^{-1}$ assuming a radiative efficiency of $0.1$ and $v_{\mathrm{SN}} = 6000$ \kms, which broadly agrees with the observed main peak luminosity. We discuss the possible mechanisms that can cause such eruptive mass loss as well as other possible progenitor scenarios in \S\ref{sec:disc:orig}.

Interestingly enough, there is a recent analogue of SN~2023aew, in terms of the precursor-like ZTF light curve (there is no TESS data to constrain the explosion in this case). SN~2023plg was discovered by ZTF and reported to TNS by ALeRCE on 2023-08-14 with a reported apparent magnitude of $18.7$. The ZTF light curve of SN~2023plg tracks closely that of SN~2023aew, both being very red during the initial decline, then brightening suddenly and maintaining a broad second peak. 
During the second peak, SN~2023plg was classified as a SN Ib on TNS by ePESSTO, but no spectroscopic data is available for the first peak. Close examination of more spectra suggests weak \ion{He}{1} lines 
similar to
what we see in SN~2023aew. The comparison is shown in Figure~\ref{fig:xuvkn}, with light curves in the upper panel and Keck1/LRIS spectra of both SNe in the lower panel. SN~2023plg is being followed up for future studies.

\subsection{Origin of SN~2023aew}\label{sec:disc:orig}
In this section, we discuss several scenarios that might explain the peculiar nature of SN~2023aew. We list several possibilities but leave a more detailed assessment and modeling of each scenario for a future study.

The precursor activities observed in SNe are often associated with mass ejection from their progenitors shortly before their explosions. One possible scenario to explain SN~2023aew is a mass ejection forming the precursor (first peak) followed by a SN explosion that shapes the major (second) light-curve peak. In the case of SN~2023aew, a hydrogen feature was observed during the precursor which weakens during the following brighter phase. Thus, the pre-SN mass ejection may have resulted in the ejection of all the remaining hydrogen-rich envelope in the progenitor and the following SN became a hydrogen-poor stripped envelope SN. Thus, the progenitor might have been similar to those of Type~IIb SNe retaining a small amount of hydrogen-rich envelope. Several stripped-envelope SNe are known to have a nearby hydrogen-rich CSM (e.g., SN~2014C, \citealt{2015ApJ...815..120M}; SN~2017ens, \citealt{P1_2017ens}). SN~2023aew could be an extreme case of a similar kind where the final ejection of the hydrogen-rich envelope occurred immediately before the explosion forming the precursor, and the following SN is observed as hydrogen-poor.

There are several suggested mechanisms that can trigger the mass ejection shortly before the explosions of massive stars. For example, an explosive nuclear shell burning may occur in the final stages of massive star evolution triggering a strong mass ejection \citep[e.g.,][]{2015ApJ...810...34W}. Mass ejection may also be triggered by an acoustic wave initiated by the strong convective motion in the core of massive stars shortly before explosion \citep[e.g.,][]{2012MNRAS.423L..92Q,2018MNRAS.476.1853F}. However, in both cases, the predicted energy that can be released is lower than the precursor energy estimated in SN~2023aew \citep[see e.g.,][]{MM2022,Wu2021,Leung2021}. If the progenitor is in a close binary system with a compact companion, the accretion to the companion may help providing additional energy to form the bright precursor \citep{Daichi2024}. 

Close passage of a massive star in an eccentric binary system could also result in precursor outbursts and was suggested for $\eta$ Car by \citet{Smith2011}. \citet{Soker2013} explored a merger-burst model (non-terminal) for SN~2009ip where the precursor outbursts were explained by close periastron passages of the massive stars, and non-spherical CSM (torus-like) is naturally expected from these encounters.

Another possibility is a pulsational pair-instability SN \citep{2007Natur.450..390W}. Pulsational instability SNe are transients caused by a partial mass ejection of very massive stars triggered by the pair instability. A pulsational mass loss forming the precursor phase can occur several months before the final collapse of the massive stars. The final collapse may result in a SN explosion forming the major light curve peak as observed in SN~2023aew. For example, some pulsational pair-instability SN models presented in \citet{2017ApJ...836..244W} have a similar luminosity to the precursor of SN~2023aew. It is possible that the progenitor of SN~2023aew experienced pulsational pair-instability mass ejection followed by a SN explosion. Although the progenitors of pulsational pair-instability SNe are expected to be massive ($\gtrsim 30~M_\odot$, e.g., \citealt{2020A&A...640A..56R}), a SN explosion with an ejecta mass of around $10~M_\odot$ could be achieved if a part of the progenitor forms a black hole. 

The ``precursor" of SN~2023aew is as bright as a typical Type~II SN. Thus, it is possible that the precursor itself is already a SN event, and the second peak is instead caused by a delayed energy injection at the center. If the hydrogen-rich layers in the ejecta are thin enough, the second peak caused by the delayed energy injection could be observed as a hydrogen-poor event. The delayed energy injection may be caused by a fallback accretion disk towards the central compact remnant \citep[e.g.,][]{2019ApJ...880...21M, Ping2023}. Depending on the initial angular momentum, the formation of the accretion disk that can provide the central energy injection could be delayed. The delayed energy input may also be caused by a delayed phase transition of neutron stars to quark stars \citep{Ouyed2013}.

\section{Summary and conclusions}\label{sec:conclusion}
In summary, SN~2023aew shows an unprecedented double-bumped light curve with two bright peaks separated by as much as  112 days. 
The light curve shares some similarities with the light curves of 09ip-like SNe with their long-lived precursors, but SN 2023aew is spectrally a stripped-envelope supernova during its main peak. SN~2023aew has a luminous 100 days long precursor, which has a 20 day rise and a spectrum similar to Type IIb SNe with a velocity of $\sim$11,800 \kms. Such a precursor is predicted by semi-analytical eruption models of \citet{MM2022} for 09ip-like SNe. After a 100 days, SN~2023aew brightens rapidly to $-18.8$ magnitude and exhibits a broader than typical (for SESNe) main peak with undulations in its decline. The bolometric light curves when fitted with the Arnett model for radioactivity power result in unreasonably high $^{56}$Ni masses, indicating the need for an additional powering source to explain the luminosity and broadness. During the main peak phase, the SN is spectrally similar to a SN Ibc, although with weaker \ion{He}{1} features and hydrogen which emerges again in the nebular phase. Photospheric and nebular phase line strengths are more similar to those of SNe Ib than to SNe Ic. Line strength of the nebular H$\alpha$/[\ion{N}{2}] complex is much higher and the line width is much smaller than expected for a normal hydrogen-free SN~Ibc, strengthening the case for hydrogen being present in the late spectra. Additionally, the nebular lines ([\ion{O}{1}, [\ion{Ca}{2}]], \ion{Mg}{1}] and H$\alpha$) show double-peaked or ``horned" profiles with one peak at rest wavelength and the other blueshifted by $\sim$1500 \kms, indicating a non-spherical geometry.

We explored the possibility of SN~2023aew being two coincident SNe in the same host galaxy and found this to be highly unlikely. We then discussed the possible origins of SN~2023aew. Although the first peak has properties consistent with a SESN (SN~IIb), the dramatic rebrightening would require either a strong delayed interaction with CSM (but no signs of this interaction are seen in the spectra) or a very delayed energy injection by a central engine. The first peak could instead be an eruptive precursor to the SN explosion (second peak), additionally powering the second peak through shock interaction but with spectral signatures of interaction hidden due to asymmetric geometry. Ultimately, the powering mechanism(s) of this double-bumped supernova remain elusive. In any case, SN~2023aew and similar SNe provide a unique opportunity to study the final throes of a dying stripped massive star and we encourage further studies with detailed theoretical modeling of the data to understand its progenitor scenario.

\section{Acknowledgement}
\small{Based on observations obtained with the Samuel Oschin Telescope 48-inch and the 60-inch Telescope at the Palomar Observatory as part of the Zwicky Transient Facility project. ZTF is supported by the National Science Foundation under Grants No. AST-2034437 and a collaboration including current partners Caltech, IPAC, the Oskar Klein Center at Stockholm University, the University of Maryland, University of California, Berkeley, the University of Wisconsin at Milwaukee, University of Warwick, Ruhr University, Cornell University, Northwestern University and Drexel University. Operations are conducted by COO, IPAC, and UW. The ZTF forced-photometry service was funded under the Heising-Simons Foundation grant \#12540303 (PI: Graham). 
  The Gordon and Betty Moore Foundation, through both the Data-Driven Investigator Program and a dedicated grant, provided critical funding for SkyPortal. The Oskar Klein Centre was funded by the Swedish Research Council. Partially based on observations made with the Nordic Optical Telescope, operated by the Nordic Optical Telescope Scientific Association at the Observatorio del Roque de los Muchachos, La Palma, Spain, of the Instituto de Astrofisica de Canarias. Some of the data presented here were obtained with ALFOSC. Some of the data presented herein were obtained at the W. M. Keck Observatory, which is operated as a scientific partnership among the California Institute of Technology, the University of California, and NASA; the observatory was made possible by the generous financial support of the W. M. Keck Foundation. The SED Machine is based upon work supported by the National Science Foundation under Grant No. 1106171. This work has made use of data from the Asteroid Terrestrial-impact Last Alert System (ATLAS) project. The Asteroid Terrestrial-impact Last Alert System (ATLAS) project is primarily funded to search for near earth asteroids through NASA grants NN12AR55G, 80NSSC18K0284, and 80NSSC18K1575; byproducts of the NEO search include images and catalogs from the survey area. The ATLAS science products have been made possible through the contributions of the University of Hawaii Institute for Astronomy, the Queen’s University Belfast, the Space Telescope Science Institute, the South African Astronomical Observatory, and The Millennium Institute of Astrophysics (MAS), Chile. This research has made use of the NASA/IPAC Infrared Science Archive, which is funded by the National Aeronautics and Space Administration and operated by the California Institute of Technology. The Liverpool Telescope is operated on the island of La Palma by Liverpool John Moores University in the Spanish Observatorio del Roque de los Muchachos of the Instituto de Astrofisica de Canarias with financial support from the UK Science and Technology Facilities Council. We acknowledge ESA Gaia, DPAC and the Photometric Science Alerts Team (http://gsaweb.ast.cam.ac.uk/alerts). Funding for the TESS mission is provided by NASA’s Science Mission directorate. This paper includes data collected by the TESS mission, which are publicly available from the Mikulski Archive for Space Telescopes (MAST). This research has made use of the Spanish Virtual Observatory (https://svo.cab.inta-csic.es) project funded by MCIN/AEI/10.13039/501100011033/ through grant PID2020-112949GB-I00. \\
Y. Sharma thanks the LSSTC Data Science Fellowship Program, which is funded by LSSTC, NSF Cybertraining Grant \#1829740, the Brinson Foundation, and the Moore Foundation; her participation in the program has benefited this work. S.~Schulze, N.~Rehemtulla and A. A. Miller are supported by LBNL Subcontract NO.~7707915. The material contained in this document is based upon work supported by a National Aeronautics and Space Administration (NASA) grant or cooperative agreement. Any opinions, findings, conclusions, or recommendations expressed in this material are those of the author and do not necessarily reflect the views of NASA. This work was supported through a NASA grant awarded to the Illinois/NASA Space Grant Consortium. \\
\texttt{Fritz} \citep{skyportal2019,duev2019real,Coughlin2023} (a dynamic collaborative platform for time-domain astronomy) was used in this work.}

\bibliography{biblio}
\bibliographystyle{aasjournal}

\appendix 
\section{Photospheric phase line strengths and velocities}\label{sec:app}

\begin{figure*}[t]
    \centering
    \includegraphics[width=0.98\textwidth]{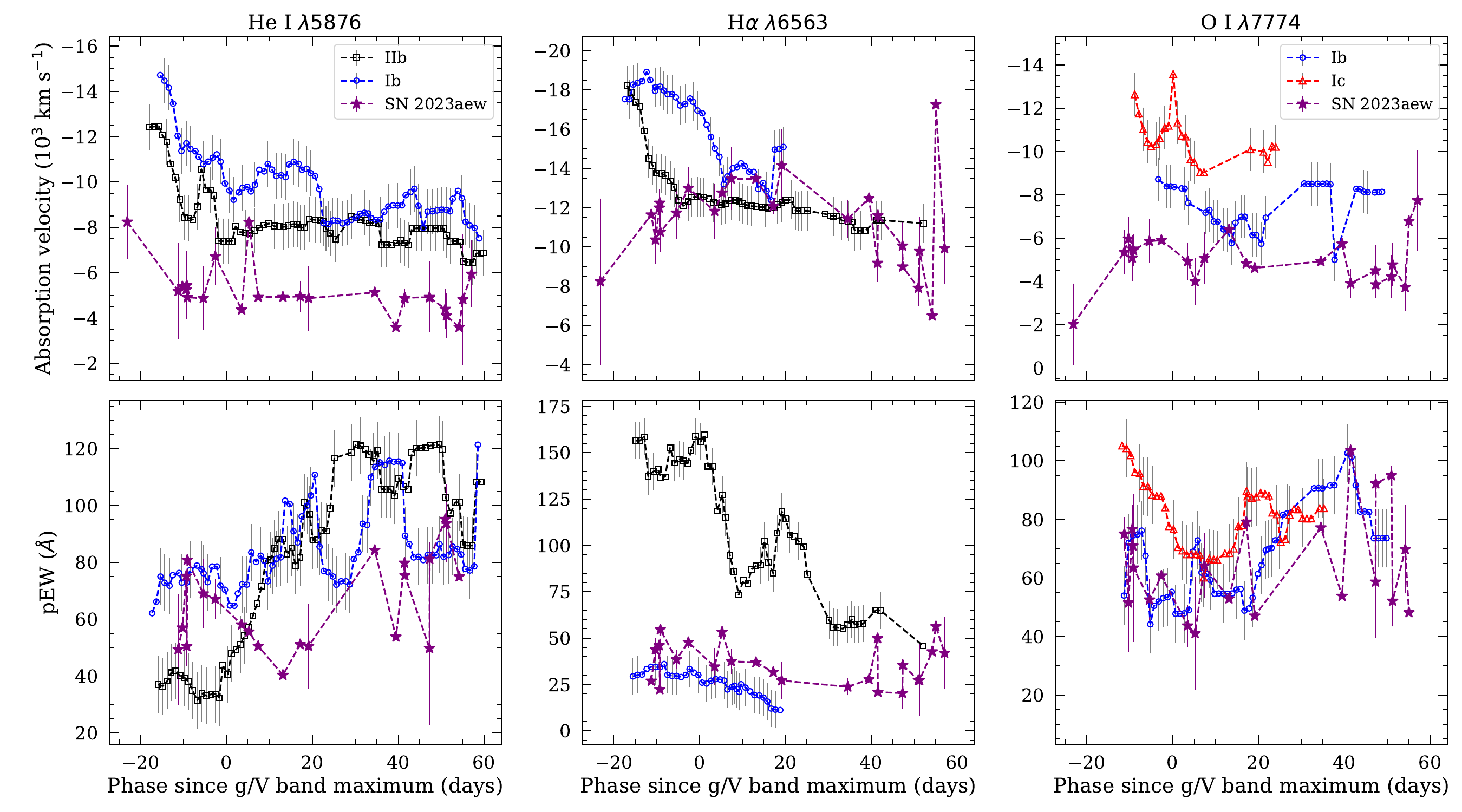}
    \caption{Absorption velocity and pseudo-equivalent width (pEW) evolution with phase for \ion{He}{1} $\lambda5876$, H$\alpha$ and \ion{O}{1} $\lambda7774$. Shown for comparison are running weighted averages of velocities and pEWs for SNe IIb, Ib and Ic with error bars of 1000\,\kms and 10\,\AA\ respectively from \citet{liu2016}. } 
    \label{fig:HeHO}
\end{figure*}

In an attempt to quantitatively compare spectral features of SN~2023aew with different SESN subtypes, the evolution of the absorption velocity and the pseudo-equivalent width (pEW) with phase were measured, and these are plotted for SN~2023aew along with mean values of these quantities for Type IIb, Ib, and Ic SNe as obtained from \citet{liu2016}. Running means of the absorption velocity and pEW during the photospheric phase ($-20$ to 60 days) for \ion{He}{1} $\lambda5876$ and H$\alpha$ for a sample of SNe IIb and Ib, and for \ion{O}{1} $\lambda7774$ for SNe Ib and Ic are shown in Figure~\ref{fig:HeHO} with values for SN~2023aew (assuming the 5876\,\AA\ feature is from helium) marked in purple stars. An average error of 1000\,\kms for velocities and 10\,\AA\ for pEW is shown instead of the exact errors from \citet{liu2016}. Similar to \citet{liu2016}, absorption velocities were estimated by fitting a smooth curve to the line absorption to find the minimum flux. For the pEWs, a local (pseudo-)continuum was estimated by fitting a low-order polynomial curve to points around maxima on either side of the absorption line which was then used to estimate the equivalent width. To estimate the uncertainty on the velocity and pEW, a median filter was applied to each spectrum and the smoothened spectrum was subtracted from the original to get residuals, then the standard deviation of these residuals was taken as the flux uncertainty at each wavelength bin of the spectrum. Assuming the flux uncertainty as the 1$\sigma$ noise which obeys a Gaussian distribution centered around the smoothened spectrum, 10000 samples were drawn to generate synthetic spectra and the line parameters were calculated on these spectra, the standard deviations of which were taken as the 1$\sigma$ errors. The phases plotted for SN~2023aew (Figure~\ref{fig:HeHO}) are not with respect to the explosion epoch but with respect to the second peak maximum (132 days from explosion) to match with the \citet{liu2016} data.

Looking at the \ion{He}{1} $\lambda5876$ and the H$\alpha$-``feature" (as marked in Figure~\ref{fig:speccomp}) absorption velocities with respect to phase in the upper left and middle panels of Figure~\ref{fig:HeHO}, SN~2023aew has lower velocities at all phases than both SNe Ib and IIb. The helium velocity is roughly constant with phase at 5000\,\kms. The trend observed in pEW (lower left and middle panels in Figure~\ref{fig:HeHO}) is more indicative of SN~2023aew being similar to SNe Ib rather than to SNe IIb for both lines. The H$\alpha$ strength is slightly more than visible in a SN Ib but much lower than for an average SN IIb, making SN~2023aew an intermediate, somewhat peculiar object between the two classes in this regard. The helium pEW is higher in SNe Ib before maximum and thereafter similar in both SNe Ib and IIb \citep{Fremling2018, liu2016}, and SN~2023aew aligns more closely with SNe Ib in Figure~\ref{fig:HeHO}. Looking at the rightmost upper and lower panels, both the \ion{O}{1} $\lambda7774$ velocity and pEW are more similar to those of SNe Ib than to SNe Ic at all epochs. Even though SN template matching programs estimate SN~2023aew to be a Type Ic at these epochs, the oxygen line strength for this SN is closer to those seen in SNe IIb/Ib and velocities are slower than for a typical SN Ic.

\section{Note added in proofs}\label{sec:Note}
At submission of this paper, another investigation of the same object was submitted to arXiv. \citet{Kangas} also present a comprehensive observational campaign on SN 2023aew, and although we do not agree on all details in the analysis there is overall agreement that this is an enigmatic unique object for which it is very difficult to determine the powering mechanism of the two peaks.

\section{Photometry tables}
\begin{table}[H]
    \centering
    \footnotesize
    \caption{Log of TESS-Red band observations of 3$\sigma$ significance (full table available online)
    \label{tab:tess}}
    \begin{tabular}{cc}
    \toprule
    \toprule
        MJD & Brightness \\
            & (mag)  \\
    \midrule
59941.177 & 19.78 $\pm$ 0.15 \\
... & \\
\bottomrule
\end{tabular}
\end{table}

\begin{table}[H]
    \centering
    \footnotesize
    \caption{Log of UVOT observations (full table available online)}
    \label{tab:app:xrt}
    \begin{tabular}{ccc}
    \toprule
    \toprule
        MJD & Filter & Brightness\\
            &        & (mag)     \\
    \midrule
60076.47	&$uvw2 $&$ 20.54 \pm 0.14 $\\
... & & \\
\bottomrule
\end{tabular}
\end{table}

\begin{table}[H]
\centering
    \caption{Log of XRT observations (full table available online)}
    \label{tab:xray}
    \begin{threeparttable}
    \begin{tabular}{cccc}
    \toprule
    \toprule
        MJD & Count rate & $F~(0.3-10~\rm keV)$ & $L~(0.3-10~\rm keV)$\\
            & ($10^{-3}~{\rm s}^{-1}$)	& ($10^{-13}~\rm erg\,s^{-1}\,cm^{-2}$)& ($10^{42}~\rm erg\,s^{-1}\,cm^{-2}$)\\
    \midrule
    \multicolumn{4}{c}{Individual epochs} \\
    \midrule
    $60076.47 \pm 0.01 $&$ <5.5	 $&$ <2.10 $&$	<0.30 $\\
    ... & & & \\
    \midrule
    \multicolumn{4}{c}{Stacking} \\
    \midrule
    $60080.97 \pm 4.51 $&$ <1.4 $&$ <0.55 $&$ <0.08 $\\
    \bottomrule
    \end{tabular}
    \begin{tablenotes}
    \centering
        \item[] The time of reference is MJD=60076.465219. The time reports the mid-time of the observation and its error indicates the extent of the time bin.
    \end{tablenotes}
    \end{threeparttable}
    \end{table}

\end{document}